\documentclass[12pt]{article}                              %% LaTeX2e
\usepackage{epsfig}
\usepackage{amsmath,amscd,amsfonts,eucal,latexsym,amssymb}
\newlength{\dinwidth}
\newlength{\dinmargin}
 \renewcommand{\theequation}{\thesection.\arabic{equation}}
 \newtheorem{Definition}{Definition}[section]
 \newtheorem{Theorem}[Definition]{Theorem}
 
 \newtheorem{Lemma}[Definition]{Lemma}

\newsymbol\rest 1316         %% restriction symbol
\def\cD{{\cal D}}

\def\cF{{\cal F}}
\def\cG{{\cal G}}
\def\cH{{\cal H}}

\def\cL{{\cal L}}
\def\cM{{\cal M}}
\def\cN{{\cal N}}

\def\cP{{\cal P}}

\def\cS{{\cal S}}

\def\cV{{\cal V}}

\def\cX{{\cal X}}

\def\bC{{\mathbb C}}

\def\bN{{\mathbb N}}

\def\bR{{\mathbb R}}

\def\a{\alpha}
\def\b{\beta}
\def\g{\gamma}        \def\G{\Gamma}
\def\d{\delta}

\def\k{\kappa}
\def\l{\lambda}       \def\L{\Lambda}

\def\o{\omega}        \def\O{\Omega}

\def\fA{{\mathfrak A}}
\def\fB{{\mathfrak B}}

\def\imply{\Rightarrow}

\def\Lpo{{\cal L}^{\uparrow}_+} % -- Lorentz group
\newsymbol\bt 1202  % \boxtimes
\newcommand{\lcrc}{\mbox{\footnotesize $\,\,\,\circ\,\,\,$}}

\newcommand{\fatnabla}{\boldsymbol{\nabla}}
\newcommand{\dirop}{{\boldsymbol{\nabla}}\!\!\!\!\!/ \,}

\newcommand{\symprod}{\circledS}
\newcommand{\Fi}{{\mathnormal{\Phi}}}
\newcommand{\FFi}{{\boldsymbol{\Phi}}}
\newcommand{\TTh}{\boldsymbol{\Theta}}
\newcommand{\TThs}{\boldsymbol{\Theta}}
\newcommand{\semidirprod}{\times} % noch zu aendern
\newcommand{\Srv}{{\cal S}(\mathbb{R}^4,V_{\rho})}
\newcommand{\hide}[1]{} % make invisible
\newcommand{\ind}{\lhd}
\newcommand{\dni}{\rhd}

\begin{document}
\noindent
\begin{center}
{ \Large \bf A Spin-Statistics Theorem for\\[6pt]
 Quantum Fields on Curved Spacetime Manifolds\\[12pt]
 in a Generally Covariant Framework}
\\[30pt]
{\large \sc   Rainer Verch}
\\[20pt]         
                 Institut f\"ur Theoretische Physik,\\
                 Universit\"at G\"ottingen,\\
                 Bunsenstr.\ 9,\\
                 D-37073 G\"ottingen, Germany\\[4pt]
                 e-mail: verch$@$theorie.physik.uni-goettingen.de
\end{center}
${}$\\[16pt]
{\small {\bf Abstract. }     
A model-independent, locally generally covariant formulation of
quantum field theory over four-dimensional, globally hyperbolic
spacetimes will be given which generalizes similar, previous
approaches.
Here, a generally covariant quantum field theory is an assignment of
quantum fields to globally hyperbolic spacetimes with spin-structure
where each quantum field propagates on the spacetime to which it is
assigned.
Imposing very natural conditions such as local general covariance,
existence of a causal dynamical law, fixed spinor- or tensor type for
all quantum fields of the theory, and that the quantum field on
Minkowski spacetime satisfies the usual conditions, it will be shown
that a spin-statistics theorem holds: If for some of the spacetimes
the corresponding quantum field obeys the ``wrong'' connection between
spin and statistics, then all quantum fields of the theory, on each
spacetime, are trivial.

 }
${}$\\[10pt]
%
%%------------------------------------------------------------------------
\section{Introduction}
\setcounter{equation}{0}
%%------------------------------------------------------------------------
%
The spin-statistics theorem of quantum field theory in Minkowski
spacetime asserts that elementary particles with integer spin must
obey Bose-statistics (``spacelike commutativity''), while those of
half-integer spin must obey Fermi-statistics (``spacelike
anti-commutativity''). Although this behaviour of elementary particles
is often taken as an experimental fact of life, it is remarkable that
in quantum field theory such a connection between two at first sight
apparently unrelated properties of particles can be deduced from a few
very basic principles: (1) Relativistic covariance, (2) stability of
matter (spectrum condition and existence of a vacuum state), (3)
localization properties of charges and (4) locality (spacelike
commutativity of observable quantities).

This deeply rooted connection between the covariance properties of
elementary particles and the behaviour under exchange of their
positions has attracted the attention of numerous researchers in
quantum field theory, and has a long history with a fair number of
general and rigorous results. Among the first are the investigations
by Pauli \cite{Pauli} and by Fierz \cite{Fierz} who proved the
spin-statistics theorem for quantum fields of arbitrary spin obeying
linear hyperbolic wave-equations in Minkowski-spacetime. The first
results on the connection between spin and
statistics in quantum field theory in a completely general,
model-independent approach (for quantum fields in the Wightman
framework) were then obtained by Burgoyne \cite{Bur} and by L\"uders
and Zumino \cite{LudZum}.  They
have subsequently been further extended and refined, particularly to
cover the situation of having several fields of different spinor types
in a quantum field theory; these theorems are presented 
in the textbooks by Jost
\cite{Jost}, by Streater and Wightman \cite{StrWi}, and by Bogoliubov,
Logunov, Todorov and Oksak \cite{BLTO}, to which we refer the reader
for further discussion and references.

The Wightman-framework takes as fundamental objects pointlike quantum
fields which may be charge-carrying and need not represent observable
quantities. The operator-algebraic approach to quantum field theory
\cite{HK,Haag} uses, instead, observable quantities as the basic
objects describing a theory of elementary particles and, at the same
time, abandons their pointlike localizability. The charge-carrying
objects and the global gauge group are, in this approach, not put in
by hand, but can be reconstructed from the observables together with
sets of states distinguished by certain localization properties
(representing the localization properties of the charges in a quantum
field theory). This is a deep result by Doplicher and Roberts
\cite{DR} arising from the profound analysis of the charge
superselection structure by Doplicher, Haag and Roberts
(see \cite{DHR,DR,Haag} and references given therein). Spin-statistics
theorems have also been derived in the 
operator-algebraic approach to quantum field theory, beginning with
works by Epstein \cite{Eps} and by Doplicher, Haag and Roberts
\cite{DHR} for the case of strictly localizable
charges. Generalizations of spin-statistics theorems to the case of
charges that can be localized in spacelike cones have been obtained by
Buchholz and Epstein \cite{BuEps}. 

A new line of development has been
introduced by the Tomita-Takesaki modular theory of von Neumann
algebras \cite{TTT} and its connection to Lorentz-transformations
which was first established in two articles by Bisognano and Wichmann
\cite{BiWi}; see the recent review by Borchers \cite{Bor} for more
information on this nowadays very important area of activity in
algebraic quantum field theory. In this context, there are
spin-statistics theorems by Guido and Longo \cite{GuiLon.spst} and by
Kuckert \cite{Kuck} in algebraic quantum field theory which take a
certain geometric action of the Tomita-Takesaki modular objects
associated with the vacuum state and distinguished algebras of quantum
field observables as the starting point.

The results just summarized concern quantum field theory on
four-dimensional Min\-kows\-ki spacetime. The present article focusses on
quantum field theory on four-dimensional curved spacetimes, but before
turning to that topic, we just mention that spin-statistics connections
have also been investigated in other settings. Among those are, in
particular, quantum field theories on flat two-dimensional spacetime
and chiral conformal quantum field theories on one-dimensional
spacetimes (e.g.\ the circle $S^1$), see e.g.\ the articles \cite{Rehr} for
the case of two dimensions and \cite{GuiLon.CSST} for chiral conformal quantum
field theory. A spin-statistics connection for so-called ``topological
geons'' has been investigated within a
diffeomorphism-covariant approach to quantum gravity \cite{DowSor,BBCT} which
is not directly related to the quantum field theoretical framework.
For the sake of completeness we mention that the spin-statistics
connection may also be violated e.g. for quantum fields having
infinitely many components; at this point we refer to \cite{BLTO} and
references cited there.

While the spin-statistics connection is well-explored in quantum field
theory on flat spacetime, offering a wealth of results, there is
little analogous to be found so far for quantum field theory on curved
spacetime manifolds. We recall that in quantum field theory on curved
spacetime one considers quantum fields propagating on a curved,
classically described spacetime background; the standard references on
that subject, from a more mathematical point of view, include
\cite{Ful,WaldII}. Clearly, the reason for lacking results on the
spin-statistics connection in curved spacetime is that the
spin-statistics theorem on Minkowski spacetime rests significantly on
Poincar\'e-covariance which possesses no counterpart in generic curved
spacetimes. In general, the isometry group of a curved spacetime will
even be trivial. Thus it is not at all clear if a spin-statistics
theorem can be established on curved spacetime in a model-independent
quantum field theoretical framework.

The situation is, of course, better when the spacetimes on which
quantum fields propagate possess still large enough isometry
groups. Such a setting has been considered recently in \cite{GLRV}. In
that article, the charge superselection theory in the
operator-algebraic approach to quantum field theory has been
generalized from the familiar case of Minkowski spacetime to
arbitrary, globally hyperbolic spacetimes. Moreover, if a
spacetime admits a spatial rotation-symmetry with isometry group 
SO(3), and also a certain time-space reflection symmetry, then a
spin-statistics theorem has been shown to hold for covariant charges,
where the spin 
is defined via the SU(2)-covering of the spatial rotation group
SO(3). A certain geometric action of Tomita-Takesaki modular objects
associated with an isometry-invariant state and distinguished algebras
of observables has been taken as input. (We refer to \cite{GLRV} for
further details and discussion.) 
Such a spin-statistics theorem applies e.g.\ for quantum field
theories on Schwarzschild-Kruskal black hole spacetimes.

However, when one is confronted with the question if there is a
connection between spin and statistics for quantum fields on general
spacetime manifolds, one finds scarcely any results. The only results
known to us have been obtained in papers by Parker and Wang
\cite{ParWg}, and by Wald \cite{Wal.S}, and they apply to the case of
quantum fields obeying linear equations of motion. The situation
considered in these two papers is, roughly speaking, as follows: A
linear quantum field propagates in the background of a (globally
hyperbolic) spacetime consisting of three regions: A ``past'' region
and a ``future'' region, both of which are isomorphic to flat
Minkowski spacetime, and an intermediate region lying between the two
(i.e. lying to the future of the ``past'' region, and to the past of
the ``future'' region) which is assumed to be non-flat. (Actually,
only particular types of spacetimes of this form are considered in
\cite{ParWg} and \cite{Wal.S}.) Then it is shown in the mentioned
articles that a quantum field of integer spin ($\le 2$) obeying a
linear wave-equation won't satisfy canonical anti-commutation
relations in the ``future'' region if canonical anti-commutation
relations were fulfilled in the ``past'' region. In other words, the
``wrong'' commutation relations are unstable under the dynamical
evolution of the quantum field in the presence of a curved spacetime
background. Likewise, a quantum field of half-integer spin ($\le 3/2$)
will no longer satisfy canonical commutation relations in the
``future'' region if it did so in the ``past'' region. It should be
noted that these results don't make reference to states (e.g., the
vacuum state in any of the flat regions), so that it is really the
non-trivial spacetime curvature in the intermediate region inducing
dynamical instability of the ``wrong'' connection between spin and
statistics at the level of the commutation relations. In that respect,
the line of argument in \cite{ParWg} and \cite{Wal.S} seems to be
restricted to free fields.  
 
Nevertheless, there are some aspects of it which are worth pointing
out since they can be generalized to model-independent quantum field
theoretical settings. So one notes that the quantum field theories in
the flat, ``past'' and ``future'' regions are ``the same'' regarding
field content and dynamics; otherwise it would be difficult to
formulate that their commutation relations are unstable under the
dynamical evolution. There is another aspect in form of the
well-posedness of the Cauchy-problem for linear fields in globally
hyperbolic spacetime, entailing that field operators located in the
``future'' are dynamically determined by the field operators located in
the ``past'' region. This property is sometimes referred to as {\it
  strong Einstein causality}, or {\it existence of a causal dynamical
  law}, and not restricted to free field theories. Thus one may
extract from the setting investigated by Parker and Wang, and by Wald,
the two following important ingredients for a quantum field theory on
curved spacetime: The parts of the theory restricted to isomorphic
spacetime regions should themselves be isomorphic (i.e., copies of
each other), and there should exist a causal dynamical law. One may
then interpret the results of \cite{ParWg} and \cite{Wal.S} as saying that,
for a certain class of curved spacetimes and for a certain class of
quantum field theories, the two said ingredients are incompatible with
assuming the ``wrong'' connection between spin and statistics.

On the basis of the mentioned ingredients, we can now abstract from
the setting of \cite{ParWg} and  \cite{Wal.S}.
We shall consider families $\{\FFi_{\bf M}\}_{{\bf M} \in \cG}$ of
quantum field theories indexed by the elements of $\cG$, the set of
all four-dimensional, globally hyperbolic spacetimes with
spin-structures ${\bf M}$. Each $\FFi_{\bf M}$ is a quantum field
propagating on the background spacetime ${\bf M}$, and it is assumed
that for each ${\bf M}$, the quantum field $\FFi_{\bf M}$ is of a
specific spinor- or tensor-type (the same for all ${\bf M}$). The
picture is that one can, for each spinor- or tensor-type, formulate
field equations that depend on the spacetime metrics in a covariant
manner. (A very simple example is $(\Box_g + m^2)\FFi_{\bf M} = 0$ for
a scalar field $\FFi_{\bf M}$ on ${\bf M} = (M,g)$, where $\Box_g$ is
the d'Alembertian associated with the metric $g$ on the
spacetime-manifold $M$.) Then there should be an isomorphism
$\alpha_{\Theta}$ between the algebras $\cF_{{\bf M}_1}(O_1)$ and
$\cF_{{\bf M}_2}(O_2)$ formed by the field operators $\FFi_{{\bf
    M}_1}(f_1)$ and $\FFi_{{\bf M}_2}(f_2)$ with ${\rm supp}\,f_j
\subset O_j$ ($j=1,2$), respectively,
\footnote{The precise mathematical sense in which the algebras are
  formed by the field operators will be explained in Sec.\ 4.  The
  $\FFi_{\bf M}$ are viewed as operator-valued distributions and the
  $f_j$ are test-spinors or test-tensors (smooth sections of compact
  support in an appropriate spinor-bundle or tensor-bundle).}
as soon as the subregions $O_j \subset {\bf M}_j$ are isomorphic,
i.e.\ whenever there is a local isomorphism (of metrics and spin-structures)
$\Theta : {\bf M}_1 \supset O_1 \to O_2 \subset {\bf M}_2$. Moreover,
$\alpha_{\Theta}$ should be a net-isomorphism in the sense that it
respects localized inclusions, meaning that
$$ \alpha_{\Theta}(\cF_{{\bf M}_1}(O)) = \cF_{{\bf M}_2}(\Theta(O)) $$
holds for all $O \subset O_1$. This is the {\it principle of general
  covariance}. It is worth noting that our concept of general
covariance is a ``local'' one, in contrast to a similar, but global
notion of general covariance for quantum field theories which has been
developed by Dimock \cite{Dim.KG,Dim.D}. Apart from that (and apart
from the fact 
that we need the net-isomorphisms at the level of von Neumann
algebras, while in existing literature they have been looked at as
$C^*$-algebraic net-isomorphisms), our concept
of general covariance is very close to that suggested by Dimock, and also
similar to ideas in \cite{Ban,Kay.RomePr,HolWald}.

The principle of existence of a causal dynamical law can then be
expressed by demanding that, for each ${\bf M}$, there holds
$$ \cF_{\bf M}(O_1) \subset \cF_{\bf M}(O) $$
whenever the subregion $O_1$ of ${\bf M}$ lies in the domain of
dependence of the subregion $O$ of ${\bf M}$ (that is, $O_1$ is
causally determined by $O$, see Sec.\ 2 for details).

There is another principle that is also most naturally
imposed. Minkowski spacetime ${\bf M}_0$ is also a member of $\cG$,
and clearly the quantum field theory $\FFi_{{\bf M}_0}$ should satisfy
the usual properties assumed for a quantum field theory (e.g., in the
Wightman framework), like Poincar\'e-covariance, spectrum condition,
existence of a vacuum state and, in order that a spin-statistics
theorem can be expected, the Bose-Fermi alternative.

If these conditions --- fixed spinor- or tensor-type, general
covariance, existence of a causal dynamical law and the usual
properties for the theory $\FFi_{{\bf M}_0}$ on Minkowski spacetime
--- are satisfied, we call the family $\{\FFi_{\bf M}\}_{{\bf M}\in
  \cG}$ a generally covariant quantum field theory over $\cG$. For
such generally covariant quantum field theories over $\cG$ we shall
establish in the present article a spin-statistics theorem. Roughly
speaking, the contents of that theorem are as follows (see Thm.\ 5.1  for
the precise statement): If there is some ${\bf M} \in \cG$ and a pair
of causally separated regions $O_1$ and $O_2$ in ${\bf M}$ so that
pairs of field operators of the quantum field $\FFi_{\bf M}$ localized
in $O_1$ and $O_2$, respectively, fulfill the ``wrong'' connection
between spin and statistics (i.e.\ they anti-commute if $\FFi_{\bf M}$
is of integer spin-type (tensorial), or they commute if $\FFi_{\bf
  M}$ is of half-integer spin type (spinorial)), then this entails
that all field operators $\FFi_{\tilde{\bf M}}$ are mutliples of the
unit operator for all $\tilde{\bf M} \in \cG$, thus the theory is
trivial. 

Our method of proof is to show  with the help of a spacetime
deformation argument (Lemma 2.1) that under the said assumptions the
``wrong'' connection between spin and statistics in any of the
theories $\FFi_{\bf M}$ leads to the ``wrong'' spin-statistics
connection for the
theory $\FFi_{{\bf M}_0}$ on Minkowski spacetime; hence the known
spin-statistics theorem for quantum field theory on Minkowski
spacetime shows that $\FFi_{{\bf M}_0}$ must be trivial. Using the
spacetime deformation argument once more, this will then be shown to
imply that all theories $\FFi_{\tilde{\bf M}}$ are trivial.

The framework we use is in a sense a mixture of the
Wightman-type quantum field theoretical setting and of the
operator-algebraic approach to quantum field
theory. This seems to have some technical advantages. Upon making some
changes, one could reformulate the arguments so that they apply
either to a purely Wightman-type quantum field theoretical setting, or
to a purely operator-algebraic approach; however in the latter case 
it wouldn't be so clear how to assign to a theory a spinor- or
tensor-type on a curved spacetime. This has resulted in the framework we
shall be employing here. 

We should like to point out that the assumptions imposed on a
generally covariant quantum field theory $\{\FFi_{\bf M}\}_{{\bf M}
  \in \cG}$ over $\cG$ are quite general. They are fulfilled for free
field theories on curved spacetimes in representations induced by
Hadamard states as we will indicate by sketching some examples in
Sec.\ 6. Our current understanding is, however, that these assumptions
aren't restricted to the case of free field theories but apply in fact
to a larger class of quantum field theories. At any rate, they reflect
a few very natural and general principles.

Our work is organized as follows.
In Sec.\ 2 we summarize a few properties of globally hyperbolic
spacetimes. Lemma 2.1 will be of importance later for proving the
spin-statistics theorem; it states that one can deform a globally
hyperbolic spacetime into another globally hyperbolic spacetime which
is partially flat, and partially isomorphic to the original
spacetime. Section 3 contains the technical definition of local
isomorphisms between spacetimes with spin structures. In Sec.\ 4 we
give the full definition of a generally covariant quantum field theory
over $\cG$. The main result on the
connection between spin and statistics for such generally covariant
quantum field theories over $\cG$ is presented in Sec.\ 5. In Sec.\ 6
we sketch the construction of three theories that provide examples for
generally covariant quantum field theories over $\cG$: The free scalar
Klein-Gordon field, the Proca field and the Majorana-Dirac field in
representations induced by quasifree Hadamard states. 

There are three appendices. Appendix A contains the proof of Lemma 2.1,
and in Appendix B we summarize the standard assumptions for a quantum
field theory on Minkowski spacetime and quote the corresponding
spin-statistics theorem from the literature. In Appendix C we briefly
indicate (generalizing similar ideas in \cite{Dim.D}) that generally
covariant quantum field theories over $\cG$ may be viewed as covariant
functors from the category $\cG$ of globally hyperbolic spacetimes
with a spin-structure to the category $\cN$ of nets of von Neumann algebras
over manifolds, both categories being equipped with suitable local
isomorphisms as morphisms.

%
%%------------------------------------------------------------------------
%
\section{Globally Hyperbolic Spacetimes}
\setcounter{equation}{0}
%
%%------------------------------------------------------------------------
%
We begin the technical discussion by collecting some basics on
globally hyperbolic spacetimes. This section will be brief, and serves
mainly for fixing our notation. The reader is referred to the
monographs \cite{HE,WaldI} for further explanations and proofs.

A spacetime is a pair $(M,g)$ where $M$ is a four-dimensional
smooth manifold (connected, Hausdorff, paracompact, without boundary)
and $g$ is 
a Lorentzian metric with signature $(+,-,-,-)$ on $M$. It will be
assumed that $(M,g)$ is orientable and time-orientable, meaning that
there exists a smooth timelike vectorfield $v$ on $M$. (Then
$g(v,v)> 0$ everywhere on $M$, so $v$ is nowhere vanishing). A
continuous, piecewise smooth causal curve $\bR \supset (a,b) \owns t
\mapsto \gamma(t)$ is future-directed (past-directed) if
$g(\dot{\g},v) > 0$ ($g(\dot{\g},v) < 0$) where $\dot{\g} =
\frac{d}{dt}\g$ is the tangent vector. Henceforth, it will be assumed
that an orientation and a time-orientation have been chosen. Then one
defines the following regions of causal dependence for any given set
$O \subset M$:
\begin{itemize}
\item[(i)] $J^\pm(O)$ is the set of all points lying on
  future(+)/past(--)\,-directed causal curves emanating from $O$,
\item[(ii)] $J(O) = J^{+}(O) \cup J^{-}(O)$,
\item[(iii)] $D^{\pm}(O)$ is the set of all points $p$ in $J^{\pm}(O)$
  such that each past(+)/future(--)\,-directed causal curve starting at
  $p$ passes through $O$ unless it has a past/future endpoint,
\item[(iv)] $D(O) = D^+(O) \cup D^-(O)$.
\item[(v)] $O^{\perp} = M \backslash \overline{J(O)}$ is the
  {\it causal complement} of $O$.
\end{itemize}
The set $D(O)$ is called the {\it domain of dependence} of $O$. If
$O_1 \subset {\rm int}\,D(O)$, then we say that $O_1$ is {\it
  causally determined} by $O$, and denote this by $O_1 \ind O$.

A time-orientable spacetime $(M,g)$ is called {\it globally
  hyperbolic} if $M$ possesses a smooth hypersurface which is
intersected exactly once by each inextendible causal curve. Such a
hypersurface is called a {\it Cauchy-surface}. It is known
that globally hyperbolic spacetimes possess
$C^{\infty}$-foliations into Cauchy-surfaces, in other words, for each
globally hyperbolic spacetime $(M,g)$ there exists a smooth
3-dimensional manifold $\Sigma_0$ together with a diffeomorphism $F: \bR
\times \Sigma_0 \to M$ such that for all $t \in \bR$, $F(\{t\} \times
\Sigma)$ is a Cauchy-surface in $(M,g)$ and such that, for each $x
\in \Sigma_0$, $\bR \owns t \mapsto F(t,x)$ is an endpointles timelike
curve. While this may at first sight appear to be
quite restrictive, it is known that the set of globally hyperbolic
spacetimes is quite large and contains many spacetimes of physical
interest. Moreover it should be noted that global hyperbolicity isn't
connected to the existence of spacetime symmetries.

When $N$ is an open, connected subset of $M$, then
$(N,g\rest N)$ is again an oriented and time-oriented spacetime. 
We call it a {\it globally hyperbolic sub-spacetime} of $(M,g)$ if the
following conditions are satisfied (cf.\ \cite{HE}{Sec.\ 6.6}): (1)
the strong causality assumption holds on $(N,g\rest N)$, (2) for
any two points $p,q \in N$, the set $J^+(p) \cap J^-(q)$, if
non-empty, is compact and contained in $N$. This entails that
$(N,g\rest N)$ is a globally hyperbolic spacetime in its own right,
but also when seen as embedded into $(M,g)$.
 We give two types of examples for subsets
$N$ of $M$ so that $(N,g\rest N)$ is a globally hyperbolic
sub-spacetime: First, if $p,q \in M$ with $p \in {\rm
  int}\,J^+({q})$, then the `double cone'
 $N = {\rm int}(J^-({p}) \cap J^+({q}))$
gives rise to a globally hyperbolic sub-spacetime. And secondly,
suppose that $C_1,C_2,C_3$ are three Cauchy-surfaces in $(M,g)$ with 
$C_2 \subset {\rm int}\,J^+(C_1)$ and $C_3 \subset {\rm
  int}\,J^+(C_2)$, and let $\cG$ be a connected open subset of $C_2$.
Then the `truncated diamond' $\cN = {\rm int}(D(\cG) \cap J^+(C_1)
\cap J^-(C_3))$ yields, equipped with the appropriate restriction of
$g$, again a globally hyperbolic sub-spacetime of $(M,g)$. 

For the purposes of the present paper, a particularly important
property of globally hyperbolic spacetimes is the following: A
globally hyperbolic spacetime $(M,g)$ can be `deformed' into another
globally hyperbolic spacetime $(\widetilde{M},\widetilde{g})$ in
such a way that certain regions of $(M,g)$ remain unchanged in
$(\widetilde{M},\widetilde{g})$, while other regions in
$(\widetilde{M},\widetilde{g})$ are isomorphic to parts of flat
Minkowski spacetime. This will be made more precise in the subsequent
statement, whose proof, given in Appendix A, is an extension of
methods used in \cite{FNW}.
\begin{Lemma}
Let $(M,g)$ be a globally hyperbolic spacetime and let $p_1,p_2 \in M$
be a pair of causally separated points (i.e.\ $p_1 \in
\{p_2\}^{\perp}$). Then there is a globally hyperbolic spacetime
$(\widetilde{M},\widetilde{g})$, together with a collection of
subsets $U_j,\widetilde{U}_j,\widehat{U}_j$ ($j = 1,2$) and
$G,\widehat{G}$, with the following properties:
\begin{itemize}
\item[(a)] There are Cauchy-surfaces $\Sigma$ in $(M,g)$, and
  $\widetilde{\Sigma}$ in $(\widetilde{M},\widetilde{g})$, so that with 
$N_+ = {\rm int}\,J^+(\Sigma) \subset M$ and $\widetilde{N}_+ =
{\rm int}\,J^+(\widetilde{\Sigma})$, $(N_+,g\rest N_+)$ is
isomorphic to
$(\widetilde{N}_+,\widetilde{g}\rest\widetilde{N}_+)$.
\item[(b)] $p_1,p_2 \in N_+$. The isomorphic images of $p_1$ and
  $p_2$ in $\widetilde{N}_+$ will be denoted by $\widetilde{p}_1$
  and $\widetilde{p}_2$.
\item[(c)] $\widehat{G} \subset \widetilde{N}_- = {\rm
    int}\,J^-(\widetilde{\Sigma})$ is simply connected, and
  $(\widetilde{G},\widetilde{g}\rest\widetilde{G})$  is a globally
  hyperbolic sub-spacetime of $(\widetilde{M},\widetilde{g})$ isomorphic
  to a globally hyperbolic sub-spacetime $(G_0,\eta\rest G_0)$ of
  flat Minkowski-spacetime $(M_0,\eta) \sim (\bR^4,{\rm
    diag}(+,-,-,-))$.
\item[(d)] $G \subset \widetilde{N}_+$ is simply connected and
$(G,\widetilde{g} \rest G)$ is a globally hyperbolic sub-spacetime of
$(\widetilde{M},\widetilde{g})$ containing $\widetilde{p}_1$ and
$\widetilde{p}_2$. 
\item[(e)] The sets $U_j,\widetilde{U}_j,\widehat{U}_j$ are, when
  equipped with the appropriate restrictions of $\widetilde{g}$ as a
  metric, globally hyperbolic, relatively compact sub-manifolds of
  $(\widetilde{M},\widetilde{g})$ which are, respectively, causally
  separated for different indices, and $\widetilde{p}_j \in U_j
  \subset G$, $\widetilde{U}_j,\widehat{U}_j \subset \widehat{G}$
  ($j = 1,2$). 
\item[(f)] $\widetilde{U}_j$ is causally determined by $U_j$, and
  $U_j$ is causally determined by $\widehat{U}_j$ ($j = 1,2$).  
\end{itemize}
\end{Lemma}
Figure 1 may help to illustrate the relations between the sets
involved in Lemma 1.\\[20pt]
\epsfig{file=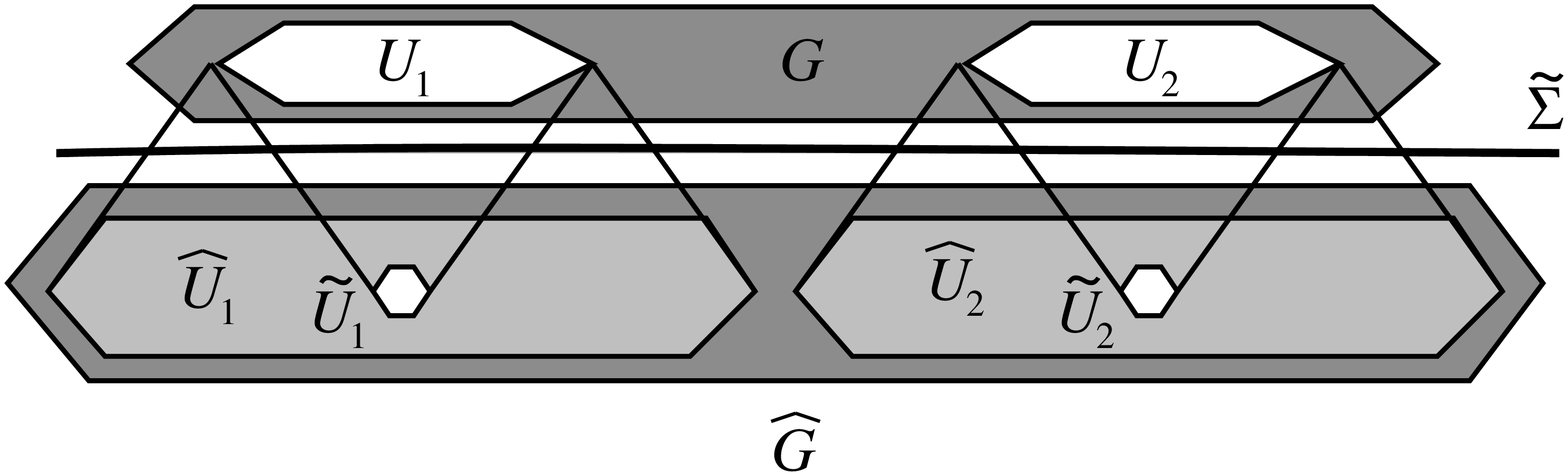, width=16cm}
${}$\begin{center}
Figure  1. Sketch of the causal relations of the sets
$U_j,\widehat{U}_j,\widetilde{U}_j$, $G,\widehat{G}$\,.
\end{center}

%
%%
%-------------------------------------------------------------
\section{Spacetimes with Spin-Structures}
\setcounter{equation}{0}
%-------------------------------------------------------------
%
%%
Let $(M,g)$ be a globally hyperbolic spacetime where an orientation
and a time-orientation have been chosen. Then let $F(M,g)$ be the
bundle of oriented and time-oriented (and future-directed)
$g$-orthonormal frames on $M$. That is, an element $e =
(e_0,\ldots,e_3)$ in $F(M,g)$ is a collection of four vectors in
$T_pM$, $p \in M$, with $g(e_a,e_b) = \eta_{ab}$ where $(\eta_{ab}) =
{\rm diag}(+,-,-,-)$ is the Minkowski metric, $e_0$ is a
future-directed timelike vector, and the frame $(e_0,\ldots,e_3)$ is
oriented according to the chosen orientation on $M$. The bundle
projection $\pi_F:F(M,g) \to M$ assigns to $e$ the base
point $p$ to which the vectors $e_0,\ldots,e_3$ are affixed. The
proper orthochronous Lorentz group $\Lpo$ operates smoothly on the right
on $F(M,g)$ by $(R_{\L}e)_a = e_b\L^b{}_a$ and thus $F(M,g)$ is a
principal fibre bundle with fibre group $\Lpo$ over $M$. A {\it spin
  structure} for $(M,g)$ is a pair $(S(M,g),\psi)$ where $S(M,g)$ is
an SL$(2,\bC)$-principal fibre bundle over $M$ and $\psi: S(M,g) \to
F(M,g)$ is a base-point preserving bundle homomorphism (that is,
$\pi_F \lcrc \psi = \pi_S$ where $\pi_S$ is the base projection of
$S(M,g)$) with the property
$$ \psi \lcrc R_{\bf s} = R_{\L({\bf s})}\lcrc \psi\,.$$
Here, $R_{\bf s}$ denotes the right action of ${\bf s} \in {\rm
  SL}(2,\bC)$ on $S(M,g)$, and SL$(2,\bC) \owns {\bf s} \mapsto
\L({\bf s}) \in \Lpo$ is the covering projection; recall that
SL$(2,\bC)$ is the universal covering group of $\Lpo$.

Two spin-structures $(S^{(1)}(M,g),\psi^{(1)})$ and
$(S^{(2)}(M,g),\psi^{(2)})$ are called (globally) {\it equivalent} if
there is a base-point preserving bundle-isomorphism
$ \Theta : S^{(1)}(M,g) \to S^{(2)}(M,g)$ so that $\Theta \lcrc
\psi^{(2)} = \psi^{(1)}$. It is known that each 4-dimensional globally
hyperbolic spacetime admits spin-structures and that all such
spin-structures are equivalent if the spacetime manifold is simply
connected (cf.\ \cite{Ger.SpSt}). 

From now on, we will abbreviate by ${\bf M} = ((M,g),S(M,g),\psi)$ an
oriented and time-oriented globally hyperbolic spacetime endowed with
a spin-structure, and we shall also use the notation ${\bf M}_j =
((M_j,g_j), S_j(M_j,g_j),\psi_j)$ if we have labels $j$ distiguishing
several such objects. We denote by ${\cal G}$ the set of all
4-dimensional, oriented and time-oriented globally hyperbolic
spacetimes with a spin-structure. One may view
$\cal G$ as a category;
of interest are then `local morphisms' between
its objects, or more properly, morphisms between sub-objects. We will
introduce the `local morphisms' as follows. For more details, see
Appendix C.
\begin{Definition}
Let ${\bf M}_1$ and ${\bf M}_2$ be in $\cG$. Then we say that
$\TTh = (\Theta,\vartheta)$ is a {\it local
  isomorphism} between ${\bf M}_1$ and ${\bf M}_2$ if:
\begin{itemize}
\item[(a)] There are simply connected, oriented and time-oriented
  globally hyperbolic sub-space\-times $(N_j,g_j\rest N_j)$ of
  $(M_j,g_j)$ $(j =1,2)$ so that $\vartheta: (N_1,g_1 \rest N_1) \to
  (N_2,g_2 \rest N_2)$ is an orientation and time-orientation
  preserving isomorphism. Then $N_1$ will be called the {\it initial
    localization} of $\TTh$, denoted by $\ell_{\rm
    ini}(\TTh)$, and $N_2$ will be called the {\it
    final localization} of $\TTh$, denoted by
  $\ell_{\rm fin}(\TTh)$.
\item[(b)] When denoting by $S_j(N_j,g_j)$ the restriction of
  $S_j(M_j,g_j)$ in its base set (that is, $S_j(N_j,g_j) =
  \pi_{S_j}^{-1}(N_j)$) , then
$$ \Theta : S_1(N_1,g_1) \to S_2(N_2,g_2) $$
is a principal fibre bundle isomorphism (so it intertwines the
corresponding right actions of the fibre groups) with the following
properties:
\begin{itemize}
\item[(i)] $\vartheta \lcrc \pi_{S_1} = \pi_{S_2} \lcrc \Theta$ \quad
  on \quad
  $S_1(N_1,g_1)$, 
\item[(ii)] $\vartheta_F\lcrc \psi_1 = \psi_2 \lcrc \Theta$ \quad on \quad
  $S_1(N_1,g_1)$.\\[2pt]
Here, $\vartheta_F : F(N_1,g_1) \to F(N_2,g_2)$ is induced by the
tangent map corresponding to $\vartheta: N_1 \to N_2$. 
\end{itemize}
\end{itemize}
\end{Definition}
{\it Remark. } In \cite{Dim.D}, Dimock has introduced the category
$\cG$, and global isomorphisms between pairs of objects in $\cG$ as
morphisms. Since each globally hyperbolic sub-spacetime of a globally
hyperbolic spacetime with spin-structure is itself a member of $\cG$,
the definition of local isomorphisms can be regarded as
introducing morphisms between sub-objects of objects in $\cG$. It
should be noted that the class of local isomorphisms between elements
of $\cG$ is clearly larger than the class of global isomorphisms as
considered in \cite{Dim.D}, and therefore covariance properties
imposed on quantum systems with respect to the class of local
isomorphisms are more restrictive than those using only global
isomorphisms. Further below we will see the implications of that.
\\[6pt]
Let $\rho$ be a linear representation of SL$(2,\bC)$ on some
finite-dimensional vector-space $V_{\rho}$ (which may be real or complex).
Then, given a spacetime-manifold with spin-structure ${\bf M}
=((M,g),S(M,g),\psi) \in \cG$, one can form the vector bundle
$$ \cV_{\rho} = S(M,g) \ltimes_{\rho} V_{\rho} $$
associated with the principal fibre bundle $S(M,g)$ and the
representation $\rho$. $\cV_{\rho}$ is a vector bundle over the base-manifold
$\cM$, and we recall that the elements of $(\cV_{\rho})_p$, the fibre
of $\cV_{\rho}$ 
at a base point $p \in M$, are the orbits $\{(R_{{\bf s}^{-1}}s_p,\rho({\bf
  s})v): {\bf s} \in {\rm SL}(2,\bC)\}$ of pairs $(s_p,v) \in S(M,g)_p
\times V_{\rho}$ under the action
\begin{equation}
\label{grpactn}
{\bf s} \mapsto (R_{{\bf s}^{-1}}s_p,\rho({\bf s})v)
\end{equation}
of the structure group SL$(2,\bC)$ of $S(M,g)$. This action induces a
linear representation $\check{\rho}$ of SL$(2,\bC)$ on each
$(\cV_{\rho})_p$. We say that $\cV_{\rho}$ is the vector bundle of (spin-)
representation type $\rho$.

Now let ${\bf M}_1$ and ${\bf M}_2$ be in $\cG$ and let $\cV_1$ and $\cV_2$
be associated vector bundles of representation type $\rho_1$ and
$\rho_2$, respectively. Suppose that $\rho_1$ and $\rho_2$ are
equivalent, i.e.\ there is some bijective linear map $T: V_1 \to V_2$
so that
\begin{equation}
\label{sprepeq}
T\rho_1(\,.\,)T^{-1} = \rho_2(\,.\,)\,.
\end{equation}
One finds from these assumptions that any local isomorphism
${\TTh} = (\Theta,\vartheta)$ between ${\bf M}_1$ and ${\bf
  M}_2$ lifts to a local isomorphism $\check{\Theta}$ between
$\cV_1$ and $\cV_2$ in a way we shall now indicate. Let
$\check{\pi}_j$ denote the base projections of $\cV_j$ $(j =1,2)$ and,
with $N_1 = \ell_{\rm ini}({\TTh})$, $N_2 = \ell_{\rm fin}({\bf
  \Theta})$, let $\cV_j(N_j) = \check{\pi}_j^{-1}(N_j)$ denote the
restrictions of the vector bundles in the base sets. Then define
$\check{\Theta}: \cV_1(N_1) \to \cV_2(N_2)$ by assigning to any
element $(s_p,v)$ in $S(M_1,g_1)_p \times V_1$, with $p \in N_1$, the
element $((\Theta s)_{\vartheta(p)},Tv)$ in
$S(M_2,g_2)_{\vartheta(p)}\times V_2$, and form the
orbits under the corresponding structure group actions
\eqref{grpactn}. It is not difficult to check that this assignment
indeed induces a well-defined map between $\cV_1(N_1)$ and
$\cV_2(N_2)$ which is linear in the fibres and fulfills
$$ \vartheta \lcrc \check{\pi}_1 = \check{\pi}_2 \lcrc
\check{\Theta}$$
on $\cV_1(N_1)$. Moreover, $\check{\Theta}$ intertwines the
representations $\check{\rho}_j$ in the sense that
$$ \check{\Theta} \lcrc \check{\rho}_1({\bf s}) = \check{\rho}_2({\bf
  s}) \lcrc \check{\Theta}$$
for all ${\bf s} \in {\rm SL}(2,\bC)$.
%
%%--------------------------------------------------------------------
\section{Generally Covariant Quantum Fields}
\setcounter{equation}{0}
%%--------------------------------------------------------------------
%
In the present section we introduce a concept of generally convariant
quantum field theories on curved spacetimes with
spin-structures. Moreover, we will make the assumption that these
quantum field theories fulfill the condition of strong Einstein
causality, or synonymously, that there exists a causal dynamical
law. The combination of these two assumptions --- general covariance
and existence of a causal dynamical law --- will lead to the
connection between spin and statistics shown in the subsequent
section.

It should be remarked that there are several possible formulations of
these two assumptions at the technical level. Here, we have chosen to
use a framework which is in a sense a mixture of the Wightman-approach
to `pointlike' quantum fields (operator-valued distributions) and the
Haag-Kastler approach which emphasizes local algebras of bounded
operators. Therefore, some technical assumptions have to be made in order
to match these two approaches; yet we feel that the resulting
framework is more general and more flexible than e.g.\ a framework
using only Wightman fields, since then we would have to make even more
stringent technical assumptions, for instance fairly detailed
assumptions on the domains of field operators, or we would have to
impose a very restrictive form of general covariance and
strong Einstein causality. Since we
don't wish to impose conditions of such kind, we regard the approach
to be presented in this section as reasonable and fairly general.

The relevant assumptions will be listed next.
\\[10pt]
{\it $(a)$ Quantum fields of a spin representation type and their
  (local) von
  Neumann algebras}\\[6pt]
Let ${\bf M} = ((M,g),S(M,g),\psi) \in \cG$ be a globally
hyperbolic spacetime with spin-structure. Moreover, let $\rho$ be a
representation of SL$(2,\bC)$ on the finite-dimensional vector-space
$V_{\rho}$. We will say that a triple
of objects $(\mathnormal{\Phi},\cD,\cH)$ is a {\it quantum field of spin
  representation type $\rho$} on ${\bf M}$ if:
$\cH$ is a Hilbert-space, $\cD$ is a dense linear subspace of $\cH$,
and $\mathnormal{\Phi}$ is a linear map taking elements $f \in
\G_0(\cV_{\rho})$, the 
space of $C^\infty$-sections in $\cV_{\rho}$ with compact support, to
closable operators $\Fi(f)$ in $\cH$ having domain $\cD$. In
addition, it will be assumed that $\cD$ is invariant under application
of the operators $\Fi(f)$, and that $\cD$ is also an invariant domain
for the adjoint field operators $\Fi(f)^*$.
It will also be assumed that there are cyclic vectors in $\cD$,
where $\chi \in \cD$ is called cyclic if the space generated by
$\chi$ and all $F_1 \cdots F_n\chi$, $n \in \bN$, where $F_j \in
\{\Fi(f_j),\Fi(f_j)^*\}$
\footnote{$\{\Fi(f_j),\Fi(f_j)^*\}$ denotes the set containing the
  operators in the curly brackets, and not their anti-commutator. In
  this work, we will never use curly brackets to denote anti-commutators.}
 with $f_j \in \G_0(\cV_{\rho})$, is dense in $\cH$.  

We write ${\rm orc}(M)$ to denote set of open, relatively compact
subsets of $M$. Let $O \in {\rm orc}(M)$, then denote by $\cF(O)$ the
von Neumann algebra which is generated by all ${\rm e}^{i\l
  |\Fi(f)|}$, $\l \in \bR$, and $J_f$, with ${\rm supp}\,f \subset
O$, where
$$ \overline{\Fi(f)} = J_f |\Fi(f)| $$
denotes the polar decomposition of a field operator's closure.
Thus the quantum field $(\Fi,\cD,\cH)$ induces a net of von Neumann
algebras $\{\cF(O)\}_{O \in {\rm orc}(M)}$ fulfilling the isotony
condition
$$ O_1 \subset O_2 \imply \cF(O_1) \subset \cF(O_2)\,.$$

In the following, we shall abbreviate a quantum field $(\Fi,\cD,\cH)$
by the symbol $\FFi$.
\\[10pt]
{\it $(b)$ Existence of a causal dynamical law}\\[6pt]
Let $\FFi$ be a quantum field of some spin-representation
type $\rho$ on ${\bf M}$. We say that there exists a {\it causal
  dynamical law} for the quantum field (or that the quantum field
fulfills {\it strong Einstein causality}) if for the net
$\{\cF(O)\}_{O \in {\rm orc}(M)}$ of local von Neumann algebras it
holds that
$$ O_1 \ind O_2 \imply \cF(O_1) \subset \cF(O_2)\,.$$
\\
{\it $(c)$ Local morphisms}\\[6pt]
Assume that we have two representations $\rho_1$ and $\rho_2$ on
finite-dimensional vector spaces $V_1$ and $V_2$, respectively,
and suppose that
these representations are isomorphic, i.e.\ \eqref{sprepeq} holds with
some bijective linear map $T: V_1 \to V_2$. Let
$\FFi_1$ and $\FFi_2$ be quantum fields of
spin-representation type $\rho_1$ and $\rho_2$ on ${\bf M}_1$ and
${\bf M}_2$, respectively, where ${\bf M}_j \in \cG$ $(j = 1,2)$. 
Moreover, suppose that there is a local isomorphism
$\TTh = (\Theta,\vartheta)$ between ${\bf M}_1$ and
${\bf M}_2$.

Then we say that
the local morphism $\TTh$ between ${\bf M}_1$ and ${\bf M}_2$ is {\it
  covered by local isomorphisms} between the quantum field theories
$\FFi_1$ and $\FFi_2$ if the following holds: Given any relatively
compact subset $N_i \subset \ell_{\rm ini}(\TTh)$ and writing $N_f
= \vartheta(N_i)$, and denoting by $\{\cF_1(O_i)\}_{O_i
  \in {\rm orc}(N_i)}$ and $\{\cF_2(O_f)\}_{O_f
  \in {\rm orc}(N_f)}$ the von Neumann algebraic nets induced by the
quantum fields $\FFi_1$ and $\FFi_2$ restricted to $N_i$ and $N_f$,
respectively, there is a
von Neumann algebraic isomorphism 
$\alpha_{\TThs,N_i}: \cF_1(N_i) \to \cF_2(N_f)$ 
fulfilling the covariance property
\begin{equation}
\label{gencov}
\alpha_{\TThs,N_i}(\cF_1(O_i)) = \cF_2(\vartheta(O_i))\,,
\quad O_i \in {\rm orc}(N_i)\,.
\end{equation}
${}$\\[6pt]
{\it Comments and Remarks.}\\[6pt]
(i) In ($a$), the property of a quantum field to be a spinor field of a
certain type is just specified by requiring that it acts linearly on
the test-spinors of the corresponding type. This is a quite common
approach to defining spinor fields on curved spacetime. An algebraic
transformation property, e.g.\ that that a (local)
spinor-transformation $\rho({\bf s})$ on $\cV_{\rho}$ induces an endomorphism
on the $*$-algebra of quantum field operators, holds in general only
when the underlying spacetime has a flat metric. One may regard the
properties of Def.\ 4.1 below as a weak replacement of such an
algebraic transformation property.
\\[6pt]
(ii) Existence of a causal dynamical law is a typical feature of
quantum fields obeying linear hyperbolic equations of motion,
but is expected to hold also for interacting quantum field theories as
long as the mass spectrum behaves moderately. For free field theories,
the existence of a causal dynamical law is commonly fulfilled in the
following stricter form (see
\cite{Dim.KG} for the case of the scalar field, but the argument
generalizes to more general types of fields, cf.\ e.g. \cite{SaV2}):
Given $O_1 \ind O_2$, then for each $f_1 \in \G_0(\cV_{\rho})$ with ${\rm
  supp}\,f_1 \subset O_1$ there is $f_2 \in \G_0(\cV_{\rho})$ with ${\rm
  supp}\,f_2 \subset O_2$ such that $\Phi(f_2) = \Phi(f_1)$. Our
formulation given in $(b)$ is more general.
\\[6pt]
(iii) It is of some importance in ($c$) that  $N_i$ and $N_f$ are
assumed to be relatively compact subsets of $\ell_{\rm ini}(\TTh)$ and
$\ell_{\rm fin}(\TTh)$, respectively, as otherwise it is known from
free field examples that a von Neumann algebraic isomorphism
$\a_{\TThs,N_i}: \cF_1(N_i) \to \cF_2(N_f)$ with the covariance
property \eqref{gencov} cannot be expected to exist. In typical cases,
the von Neumann algebras $\cF_j(O)$ are of properly infinite type,
and then $\alpha_{\TThs,N_i}$ is implemented by a unitary
operator $U_{\TThs,N_i} 
:\cH_1 \to \cH_2$. 
\\[10pt]
The subsequent definition will fix the notion of general covariance
for quantum fields on curved spacetimes.
\begin{Definition}
Let $\rho$ be a linear representation of SL$(2,\bC)$ on a finite
dimensional vector space $V$. By
$\cG$ we denote, as before, the set of all oriented and time-oriented,
4-dimensional, globally hyperbolic spacetimes equipped with a
spin-structure.
A family $\{\FFi_{\bf M}\}_{{\bf M} \in
  \cG}$ will be called a {\em generally covariant quantum field theory
  over} $\cG$ {\it of spin representation type} $\rho$ if the following
properties are fulfilled:
\begin{itemize}
\item[{\rm (A)}] 
For each ${\bf M} \in \cG$, $\FFi_{\bf M}= (\Fi_{\bf M},\cD_{\bf
    M},\cH_{\bf M})$ is a quantum field theory on ${\bf M}$ of spin
  representation type $\rho$ (the same for all ${\bf M}$) such that
  the properties (a) and (b) stated above are satisfied.
\item[{\rm (B)}] For the case that ${\bf M} = {\bf M}_0$ is Minkowski
  spacetime with its usual spin-structure, we demand that the
  corresponding quantum field theory $\FFi_{{\bf M}_0}$
 fulfills the Wightman axioms, including
  the Bose-Fermi alternative (or normal commutation relations); see
  Appendix B for details.
\item[{\rm (C)}] If for a pair ${\bf M}_1$ and ${\bf M}_2$ in $\cG$ there is
  a local isomorphism $\TTh$ 
  between $M_1$ and $M_2$, then it is covered by local isomorphisms between the
  corresponding quantum field theories $\FFi_{{\bf M}_1}$
   and  $\FFi_{{\bf M}_2}$.
\end{itemize}
\end{Definition}
Let us discuss some features of that definition in a further set of
\\[10pt]
{\it Comments and Remarks.}\\[6pt]
(iv) Readers familiar with the articles  of Dimock \cite{Dim.KG,Dim.D}
will notice that our definition is very much inspired by the concept
of general covariance introduced in those works for quantum field
theories on curved spacetimes. The main difference, as we have
mentioned already in the Remark below Def.\ 3.1, is that the
isomorphisms between the spacetimes with spin-structures, and
accordingly between the corresponding quantum field theories, are here
assumed to be local, whereas in \cite{Dim.KG,Dim.D} they are
assumed to be global. To allow local isomorphisms in the condition of
general covariance (C) leads, in combination with the conditions (A)
and (B), to restrictions which apparently are not present when using
only global isomorphisms.

The significance of that point has, in a somewhat different context,
been noted by Kay \cite{Kay.RomePr}. Our definition of a
generally covariant quantum field theory resembles an approach taken
by Kay in his investigation of ``F-locality'' in
\cite{Kay.RomePr}. The main difference
(apart from differences of technical detail) is that Kay considers a
much larger class $\widehat{\cG}$ of spacetimes which need not be
globally hyperbolic, and he essentially investigates the question of
what the largest class $\widehat{\cG}$ of spacetimes might be so that
a quantum field theory over $\widehat{\cG}$ is compatible with the
covariance property (C) once certain properties are assumed for the
quantum fields on the individual spacetimes in $\widehat{\cG}$. 
For the case of the
scalar Klein-Gordon field, he finds that 
restrictions on the class of spacetimes $\widehat{\cG}$ arise in order
to obtain compatibility, see \cite{Kay.RomePr} for further
discussion.
\\[6pt]
(v) Given a local isomorphism $\TTh$ between ${\bf
  M}_1$  and ${\bf M}_2$ in $\cG$, then it is known for free fields that
typically the identification
$$ \Fi_{{\bf M}_2} \lcrc \check{\Theta}^{\star}(f) = \Fi_{{\bf M}_1}(f)\,,
\quad {\rm supp}\, f \in \ell_{\rm ini}(\TTh)\,,\quad {\rm with} \ \
\check{\Theta}^{\star}f = \check{\Theta}\lcrc f \lcrc \vartheta^{-1}\,,$$
preserves CAR or CCR and thereby
gives rise to a ($C^*$-algebraic) local isomorphism
$\alpha_{\TThs}$ covering $\TTh$ between
the quantum field theories.
In \cite{WaldII} (pp.\ 89-91 of that reference), such a covariance
property has been proposed as a condition on the (renormalized)
stress-energy tensor of a quantum field on curved spacetimes, and more
recently, Hollands and Wald have defined the notion of a local,
covariant quantum field by means of such a covariance behaviour of the
quantum field and have shown that one may construct, essentially
uniquely, Wick-polynomials
of the free scalar field in such a way that they become local,
covariant quantum fields \cite{HolWald}.
 Our conditions on a local isomorphism
between quantum field theories are much less detailed; indeed, the
slightly complicated definition of a local isomorphism between quantum
field theories serves the purpose of keeping this notion as general as
possible and yet to transfer enough algebraic information for making
it a useful (i.e.\ sufficiently restrictive) concept in combination
with existence of a causal dynamical law formulated in (b). 
\\[6pt]
(vi) We have required that the spin-representation $\rho$ be the same
for all members $\FFi_{\bf M}$ of a
generally covariant quantum field theory over $\cG$, expressing that
all these quantum field theories on the various spacetime have the
same field content. (Of course, it would be sufficient just to require
that the various $\rho_{\bf M}$ be isomorphic; to demand equality is
just a simplification of notation.) We think that this is necessary in
order that (C) can be fulfilled, but a proof of that remains to be
given.
\\[6pt]
(vii) It should be noted that each element ${\bf M} \in \cG$ comes
equipped with an orientation and a time-orientation. The local
isomorphisms have been assumed to preserve orientation and
time-orientation, so the condition of general covariance imposes no
restrictions on quantum field theories $\FFi_{{\bf M}_1}$ and
$\FFi_{{\bf M}_2}$ when ${\bf M}_1$ and ${\bf M}_2$ are connected by a
local isomorphism that reverses orientation and time-orientation. In
fact, if $\TTh$ is an (appropriately defined)
local isomorphism between ${\bf M}_1$ and ${\bf
  M}_2$ reversing both time-orientation and orientation, one would
expect that for any relatively compact $N_i \subset \ell_{\rm
  ini}(\TTh)$, writing $N_f = \vartheta(N_i)$, there is an {\it
  anti-linear} von Neumann algebraic isomorphism $\a_{\TTh,N_i}$ 
having the covariance property (4.1). It would be quite interesting to
see if one could deduce the existence of such anti-linear local von
Neumann algebraic isomorphisms at least for a distinguished class of
time-orientation and orientation reversing local isomorphisms $\TTh$
from the assumptions on $\{\FFi_{\bf M}\}_{{\bf M} \in \cG}$ of Def.\
4.1. That would correspond to a PCT-theorem in the present general
setting. 
\\[6pt] (viii) The assignment of quantum field theories $\FFi_{\bf M}$
to each ${\bf M} \in \cG$ fulfilling the condition of general
covariance allows a functorial description which will be indicated in
Appendix C.
%

%--------------------------------------------------------------------
\section{Spin and Statistics}
\setcounter{equation}{0}
%--------------------------------------------------------------------
%
%%
In the present section we state and prove a spin-statistics theorem
for generally covariant quantum field theories over $\cG$. Before we
can start to formulate the result, it is in order to briefly
recapitulate the terminology referring to ``integer'' and
``half-integer'' spin.

Let $\symprod^k \bC^2$ denote the $k$-fold symmetrized tensor product
of $\bC^2$. Then an irreducible complex linear representation $D^{(k,l)}$
of SL$(2,\bC)$ for $k,l \in \bN_0$ is given on the vectorspace
$V_{k,l} = (\symprod^k \bC^2) \otimes (\symprod^l \bC^2)$ by
$$ D^{(k,l)}({\bf s}) = (\symprod^k {\bf s}) \otimes (\symprod^l
\overline{\bf s}) $$
where ${\bf s} \in {\rm SL}(2,\bC)$ acts like a
matrix on column vectors in $\bC^2$, and $\overline{\bf s}$ is the
matrix with complex conjugate entries.
\footnote{By convention, the case $k = 0$ and $l = 0$ corresponds to a
  scalar field, with the trivial one-dimensional representation of
  SL$(2,\bC)$.} 
 All finite-dimensional complex linear
irreducible representations of SL$(2,\bC)$ arise in this way.
Such  an irreducible representation 
 is said to be of {\it integer type} (or simply
 {\it integer}) if $k + l$ is even and of {\it half-integer type} (or
 simply {\it half-integer}) if $k +l$ is odd. There also the (finite
 dimensional) real linear irreducible representations $D^{(k,l)}
 \oplus D^{(l,k)}$ for $k \ne l$, and $D^{(l,l)}$. They are called
 real-linear irreducible because it is possible to select 
 real-linear subspaces in $V_{k,l} \oplus V_{l,k}$ and in $V_{l,l}$,
 respectively, on which these representations act irreducibly as
 real-linear representations. As complex linear representations they
 are, however, reducible except for the case $D^{(l,l)}$. The
 classification of these representations as being of ``integer'' or
 ``half-integer'' type is analogous to that of complex linear
 irreducible representations.
\begin{Theorem}
Let $\{\FFi_{\bf M}\}_{{\bf M} \in \cG}$ be a
generally covariant quantum field theory over $\cG$ of spin
representation type $\rho$, where $\rho$ is assumed to be a complex
linear irreducible, or real linear irreducible, finite dimensional
representation of ${\rm SL}(2,\bC)$. 
\\[6pt]
{\rm (I)} If $\rho$ is of half-integer type,
 and if there exist an ${\bf M} \in \cG$
and a pair of non-empty
 $O_1,O_2 \in {\rm orc}(M)$ with $O_1 \subset O_2^{\perp}$
so that $\cF_{\bf M}(O_1) \subset \cF_{\bf M}(O_2)'$\ \,\, (where by $\cF_{\bf
  M}(O)$ we denote the local von Neumann algebras generated by
$\FFi_{\bf M}$ and by $\cF_{\bf M}(O)'$ the commutant algebras\footnote{i.e.\ $\cF_{\bf M}(O)' = \{A' \in B(\cH_{\bf M}): A'A =
  AA'\,, \ \forall\ A \in \cF_{\bf M}(O)\}$\,.})
 then it follows for
all $\hat{\bf M} \in \cG$ that $\Fi_{\hat{\bf M}}(f) = c_f\cdot 1$
for some $c_f \in \bC$, i.e.\ the quantum field operators 
of all quantum fields of the
generally covariant theory are multiples of the unit operator. 
\\[6pt]
{\rm (II)} If $\rho$ is of integer type, 
and if there exist an ${\bf M} \in \cG$, a
pair of causally separated points $p_1$ and $p_2$ in $M$ and for 
each pair of open neighbourhoods $O_j$ of $p_j$ with $O_1 \subset
O_2^{\perp}$ a pair
$f_j \in \G_0(\cV_{\rho})$ with ${\rm supp}\,f_j \subset O_j$ and $\Fi_{\bf
  M}(f_j) \ne 0$ $(j = 1,2)$ so that
\begin{equation}
\label{wrongcr}
 \Fi_{\bf M}(f_1)\Fi_{\bf M}(f_2) + \Fi_{\bf M}(f_2)\Fi_{\bf
  M}(f_1) = 0\ \ {\it or}\ \ \Fi_{\bf M}(f_1)\Fi_{\bf M}(f_2)^* +
\Fi_{\bf M}(f_2)^*\Fi_{\bf 
  M}(f_1) = 0\,,
\end{equation} 
then it follows again for all $\hat{\bf M} \in \cG$ that all field
operators $\Fi_{\hat{\bf M}}(f)$ are multiples of the unit operator.
\end{Theorem}
We note that $\cF_{\bf M}(O_1) \subset \cF_{\bf M}(O_2)'$ means that
the field operators $\Phi_{\bf M}(f_1)$ and $\Phi_{\bf M}(f_2)$ for
${\rm supp}\,f_j \subset O_j$ commute strongly in the sense that the
operators appearing in their polar decompositions commute
strongly. This stronger form of commutativity at causal separation is
expected to hold in physically relevant theories. In Appendix B we
give a few more comments on this point.
 If the stronger forms of general covariance at the level of invidual field
operators as indicated in Remarks (ii) and
(iv) of Sec.\ 4 were assumed, the statement for the half-integer case could be
strengthened to resemble the integer case more closely; namely, then
one would conclude for the half-integer case 
that the relations $\Fi_{\bf M}(f_1)\Fi_{\bf
  M}(f_2) - \Fi_{\bf M}(f_2)\Fi_{\bf M}(f_1) = 0$ or
$\Fi_{\bf M}(f_1)\Fi_{\bf
  M}(f_2)^* - \Fi_{\bf M}(f_2)^*\Fi_{\bf M}(f_1) = 0$ for some ${\bf M}$
and a pair
test-spinors $f_1$ and $f_2$ with causally separated supports so that
$\Fi_{\bf M}(f_j) \ne 0$ already implies that the field operators
$\Fi_{\hat{\bf M}}(f)$ are multiples of unity for all $\hat{\bf M}
\in \cG$.
\\[6pt]
{\it Proof of Theorem 4.1. }
We begin with part (I) of the statement involving a theory of
half-integer type, and we suppose that $\cF(O_1) \subset \cF(O_2)'$
for a pair of causally separated $O_1,O_2 \in {\rm orc}(M)$, where we
use the notation $\cF(O) = \cF_{\bf M}(O)$. Then let $p_j \in O_j$,
and choose for this pair of causally separated points in $M$ a
globally hyperbolic spacetime $(\widetilde{M},\widetilde{g})$ with
neighbourhoods $U_j,\widehat{U}_j,\widetilde{U}_j$, $G,\widehat{G}$,
as in Lemma 2.1, which can be done in such a way that
$\vartheta^{-1}(U_j) \subset O_j$, where $\vartheta$ is the
isomorphism $M \supset N \to \widetilde{N} \subset \widetilde{M}$.
Now we equip $(\widetilde{M},\widetilde{g})$ with any spin-structure
and denote the resulting spacetime with spin-structure by
$\widetilde{\bf M}$. The
neighbourhoods $G$ and $\widehat{G}$ are simply connected. Thus,
since all spin-structures over simply connected globally hyperbolic
spacetimes are equivalent, there is a local isomorphism
$\TTh$ between ${\bf M}$ and $\widetilde{\bf M}$ with
$\ell_{\rm fin}(\TTh) = G$, and also a local
isomorphism $\TTh_0$ between $\widetilde{\bf M}$ and
$\widetilde{\bf M}_0$ where $\widetilde{\bf M}_0$ is Minkowski
spacetime with its standard spin-structure. This is due to the fact
that $G$ is isomorphic to a subset $\vartheta^{-1}(G)$ in $M$ and
$\widehat{G}$ is isomorphic to a subset in Minkowski-spacetime
$M_0$, cf.\ Lemma 2.1. 

Let us now introduce the notation $\widetilde{\cF}(U) =
\cF_{\widetilde{\bf M}}(U)$ and $\cF_0(U) = \cF_{{\bf M}_0}(U)$ for
the local von Neumann algebras corresponding to the theories
$\FFi_{\widetilde{{\bf M}}}$ and $\FFi_{{\bf M}_0}$, respectively. 
Then choose two globally hyperbolic, relatively compact submanifolds $N_f$ and
$\widehat{N}_i$ of $G$ and $\widehat{G}$, respectively, with the
additional property that $U_j \subset N_f$ and
$\widetilde{U}_j,\widehat{U}_j \subset \widehat{N}_i$
$(j=1,2)$. Denote $N_i = \vartheta^{-1}(N_f)$.
According
to the general covariance assumption (C) there are local isomorphisms
$\a_{\TThs,N_i}$ between $\FFi_{\bf M}$ and
$\FFi_{\widetilde{\bf M}}$ and $\a_{{\TTh}_0,\widehat{N}_i}$ between
$\FFi_{\widetilde{\bf M}}$ and $\FFi_{{\bf M}_0}$ so that
\begin{eqnarray}
\a_{\TThs,N_i}(\cF(\vartheta^{-1}(U))) &=&
\widetilde{\cF}(U)\,, \quad U \in {\rm orc}(N_f)\,, \label{5.ii}  \\
\a_{\TThs_0,\widehat{N}_i}(\widetilde{\cF}(\widehat{U})) &=&
\cF_0(\vartheta_0(\widehat{U}))\,, \quad \widehat{U} \in {\rm
  orc}(\widehat{N}_i)\,, \label{5.iii}
\end{eqnarray}
where $\vartheta_0$ is the isomorphism embedding $\widehat{G}$ into
$M_0$.
Since we have supposed initially that $\cF(O_1) \subset \cF(O_2)'$,
and since $\vartheta^{-1}(U_j) \subset O_j$, relation \eqref{5.ii}
 implies that
$\widetilde{\cF}(U_1) \subset \widetilde{\cF}(U_2)'$. Moreover, $U_j
\dni \widetilde{U}_j$ and hence, by the existence of a causal
dynamical law, it follows that 
$$ \widetilde{\cF}(\widetilde{U}_1) \subset
\widetilde{\cF}(\widetilde{U}_2)'\,.$$
Exploiting also \eqref{5.iii}, one obtains 
\begin{equation}
\label{square}
\cF_0(\vartheta_0(\widetilde{U}_1)) \subset
\cF_0(\vartheta_0(\widetilde{U}_2))' 
\end{equation}
where $\vartheta_0(\widetilde{U}_1)$ and
$\vartheta_0(\widetilde{U}_2)$ are a pair of open, causally separated
subsets of Minkowski spacetime. Since the quantum field theory
$\FFi_{{\bf M}_0}$ on Minkowski spacetime has been assumed to fulfill
the usual assumptions, and is, by assumption, of half-integer
spin-type, the last relation \eqref{square} implies by the known
spin-statistics theorem for quantum field theories on Minkowski
spacetime that $\cF_0(U_0) = \bC\cdot 1$ holds for all $U_0 \in {\rm
  orc}(M_0)$. (See Appendix B for details.)

In a next step we will show how that conclusion implies that all other
quantum field theories $\FFi_{\hat{\bf M}}$ are likewise trivial. Let
$\hat{\bf M} = ((\hat{M},\hat{g}),S(\hat{M},\hat{g}),\hat{\psi}) \in
\cG$ and choose any point $p_1 \in \hat{M}$ (and any other causally
separated point $p_2 \in \hat{M}$, which actually plays no role). Then
choose a spacetime $(\widetilde{M},\widetilde{g})$ with subsets
$U_j,\widetilde{U}_j,\widehat{U}_j$, $G,\widehat{G}$ as in Lemma 2.1 for
these data, $(\hat{M},\hat{g})$ now playing the role of
$(M,g)$. Identifying $\cF(O) = \cF_{\hat{\bf M}}(O)$ and making
similar adaptations, equations \eqref{5.ii} and \eqref{5.iii}
 hold accordingly. Then
$\cF_0(\vartheta_0(\widehat{U}_1)) = \bC \cdot 1$ implies, by \eqref{5.iii},
$\widetilde{\cF}(\widehat{U}_1) = \bC \cdot 1$, and since
$\widehat{U}_1 \dni U_1$ it follows that
$\widetilde{\cF}({U}_1) = \bC \cdot 1$. Hence \eqref{5.ii} leads to
$\cF(\vartheta^{-1}({U}_1)) = \bC \cdot 1$, implying that
$\Fi_{\hat{M}}(f)$ is a multiple of the unit operator for all $f$ with
${\rm supp}\,f \subset \vartheta^{-1}({U}_1)$.  As
$\vartheta^{-1}({U}_1)$ is an open neighbourhood of an
arbitrary point $p_1 \in \hat{M}$, and since the quantum field $f
\mapsto \Fi_{\hat{\bf M}}(f)$ is linear, a partition of unity argument
shows that therefore one must have $\Fi_{\hat{\bf M}}(f) = c_f \cdot 1$
with suitable $c_f \in \bC$ for all test-spinors $f$ on $\hat{M}$.

Now we turn to the proof of statement (II) of the theorem. According
to the assumptions, there are two points $p_1$ and $p_2$ in $M$ which
are causally separated, and moreover, when choosing a deformation
$(\widetilde{M},\widetilde{g})$ of $(M,g)$ with neighbourhoods
$U_j,\widetilde{U}_j,\widehat{U}_j$, $G,\widehat{G}$ as in Lemma 2.1,
there are a pair of testing spinors $f_j$ supported in
$\vartheta^{-1}(U_j)$ so that $\Fi_{\bf M}(f_j) \ne 0$ and such that
one of the relations \eqref{wrongcr} holds. We shall, for the sake of
simplicity of notation, assume that 
\begin{equation}
\label{anticcr}
\Fi_{\bf M}(f_1)\Fi_{\bf M}(f_2) + \Fi_{\bf M}(f_2)\Fi_{\bf M}(f_1) =
0
\end{equation}
holds, and
we will show that these properties are in conflict with Bosonic
commutation relations for the theory $\FFi_{{\bf M}_0}$ on Minkowski
spacetime. The other case of \eqref{wrongcr} can be treated by similar
arguments. The proof proceeds indirectly, so we suppose that
$\FFi_{{\bf M}_0}$ possesses Bosonic commutation relations. As before
in the proof of (I) above, we can find local isomorphisms
$\a_{\boldsymbol{\Theta},N_i}$ and
$\a_{\boldsymbol{\Theta}_0,\widehat{N}_i}$
fulfilling the relations \eqref{5.ii} and \eqref{5.iii} for the von
Neumann algebraic nets 
corresponding to the quantum field theories on ${\bf M}$,
$\widetilde{\bf M}$ and ${\bf M}_0$. Having supposed Bosonic
commutation relations for the quantum field theory on Minkowski
spacetime, it follows by \eqref{5.iii} that $\widetilde{\cF}(\widehat{U}_1)
\subset \widetilde{\cF}(\widehat{U}_2)'$. Now $U_j \ind \widehat{U}_j$
and thus, by the existence of a causal dynamical law, it holds that
$\widetilde{\cF}(U_1) \subset \widetilde{\cF}(U_2)'$. By \eqref{5.ii} we obtain
$\cF(\vartheta^{-1}(U_1)) \subset \cF(\vartheta^{-1}(U_2))'$. Since
the operators $\Fi_{\bf M}(f_j)$ are affiliated to the von Neumann
algebras $\cF(\vartheta^{-1}(U_j))$, one concludes that
\begin{equation}
\label{ccr}
\Fi_{\bf M}(f_1)\Fi_{\bf M}(f_2) - \Fi_{\bf M}(f_2)\Fi_{\bf M}(f_1) =
0\,.
\end{equation}
Comparing \eqref{anticcr} and \eqref{ccr} yields
$$      \Fi_{\bf M}(f_1) \Fi_{\bf M}(f_2) = 0 \,.$$
It is clear that this relation entails 
$\Fi_{\bf M}(f_1)^*\Fi_{\bf M}(f_1)\Fi_{\bf
  M}(f_2)\Fi_{\bf M}(f_2)^* = 0$\,. The operators $A_1 = \Fi_{\bf
  M}(f_1)^*\Fi_{\bf M}(f_1)$ and $A_2 = \Fi_{\bf M}(f_2)\Fi_{\bf
  M}(f_2)^*$ are positive and possess selfadjoint extensions
affiliated to $\cF(\vartheta^{-1}(U_1))$ and
$\cF(\vartheta^{-1}(U_2))$, respectively. Denoting by $E_j(a)$ their
spectral projections corresponding to the spectral interval $(-a,a)$,
the operators $A_j(a) = E_j(a)A_j$ are contained in
$\cF(\vartheta^{-1}(\widehat{U}_1))$ and it holds that $A_1(a)A_2(a)
= 0$ for all $a > 0$. Repeating the arguments that led to eq.\
\eqref{ccr}, one can see that the $A_j(a)$ possess isomorphic images
$\widehat{A}_j(a)$ in $\cF_0(\vartheta_0(\widehat{U}_j))$ so that
$\widehat{A}_1(a)\widehat{A}_2(a) = 0$ for all $a > 0$. But since the
net $\{\cF_0(U)\}_{U \in {\rm orc}(M_0)}$ was assumed to fulfill
Bosonic commutations relations, and since it fulfills the usual
assumptions for a quantum field theory on Minkowski-spacetime,
including spectrum condition and the existence of a vacuum state, it
follows that the Schlieder property \cite{Schl} holds for this
net. This property states that the relations $\widehat{A}_j(a) \in
\cF_0(\vartheta_0(\widehat{U}_j))$,
 ${\rm cl}\,\vartheta_0(\widehat{U}_1) \subset
 \vartheta_0(\widehat{U}_2){}^{\perp}$ 
and $\widehat{A}_1(a)\widehat{A}_2(a) = 0$ imply $\widehat{A}_1(a) =
0$ or $\widehat{A}_2(a) = 0$. Hence one obtains that, for all $a > 0$,
$A_1(a) = 0$ or $A_2(a)=0$, and this entails $A_1 = 0$ or $A_2 = 0$,
which in turn enforces $\Fi_{\bf M}(f_1) = 0$ or $\Fi_{\bf M}(f_2) =
0$. Thus one arrives at a contradiction since both operators $\Fi_{\bf
  M}(f_1)$ and $\Fi_{\bf M}(f_2)$ are by assumption different from
0. One concludes that Bosonic commutation relations are an impossible
option for the theory $\FFi_{{\bf M}_0}$ on Minkowski spacetime and
thus, due to the Bose-Fermi-alternative, that theory must fulfill
Fermionic commutation relations. Since the theory is of integer
spin-type, this implies that the von Neumann algebras $\cF_0(U_0)$ of
the theory on Minkowski spacetime consist only of multiples of the
unit operator because of the spin-statistics theorem on flat spacetime
(cf.\ Appendix B).
Repeating the argument given for part (I) above, it follows that
for each $\hat{\bf M} \in \cG$ the quantum field operators
$\Fi_{\hat{\bf M}}(f)$ are multiples of the unit operator for all
test-tensors $f$.${}$\quad ${}$ \hfill $\Box$ 
%%
%
%---------------------------------------------------------------------
\section{Examples}
%---------------------------------------------------------------------
\setcounter{equation}{0}
%---------------------------------------------------------------------
%
%
In this section we briefly indicate examples of linear quantum field
theories which fulfill the properties required for a generally
covariant quantum field theory over $\cG$ in Section 4.
\\[10pt]
{\it 1. The free scalar field. } The simplest example is the free
scalar field, although its significance for a spin-statistics theorem
is, naturally, quite limited.

For each globally hyperbolic spacetime ${\bf M} = (M,g) \in \cG$
(endowed with a spin-structure whose explicit appearance is now
suppressed since it is irrelevant for the scalar field) we consider
the scalar Klein-Gordon equation
$$ (\Box_g + m^2)\varphi = 0 $$
for real-valued functions $\varphi$ on $M$, where $m \ge 0$ is a
constant independent of ${\bf M}$ and $\Box_g$ is the scalar
d'Alembertain for $(M,g)$. Following Dimock \cite{Dim.KG}, one can
construct a $C^*$-algebraic quantization of this field as
follows. There are uniquely determined, continuous linear maps $E_{\bf
  M}^{\pm}: C_0^{\infty}(M,\bR) \to C^{\infty}(M,\bR)$ with the
properties
$$ (\Box_g + m^2)E^{\pm}_{\bf M} = f = E^{\pm}_{\bf M}(\Box_g + m^2)f\
\ {\rm and}\ \ {\rm supp}\,E^{\pm}_{\bf M}f \subset J^{\pm}({\rm
  supp}\,f)\,, \quad f \in C_0^{\infty}(M,\bR)\,.$$
Their difference $E^{\pm}_{\bf M} = E^+_{\bf M} - E^-_{\bf M}$, called
the (causal) propagator, induces a symplectic form
$$ \k_{\bf M}([f],[h]) = \int_M d\eta\,f \cdot Eh\,, \quad [f],[h] \in
K_{\bf M}\,,$$
on $K_{\bf M} = C_0^{\infty}(M,\bR)/{\rm ker}\,E_{\bf M}$, where $f
\mapsto [f] = [f]_{\bf M}$ denotes the quotient map and $d\eta$ is the
metric-induced volume-form on $(M,g)$. To the resulting symplectic
space $(K_{\bf M},\k_{\bf M})$ there corresponds the CCR-Weyl algebra
$\fA[K_{\bf M},\k_{\bf M}]$, defined as the (up to $C^*$-isomorphisms
unique) $C^*$-algebra generated by unitary elements $W_{\bf M}(x)$, $x
\in K_{\bf M}$, fullfilling the Weyl-relations, or ``exponentiated''
canonical commutation relations (see \cite{BR2})
$$ W_{\bf M}(x)W_{\bf M}(y) = {\rm exp}(-i\k_{\bf M}(x,y)/2)W_{\bf M}(x
+ y)\,, \ \ W_{\bf M}(x)^* = W_{\bf M}(-x)\,, \quad x,y \in K_{\bf M}\,.$$
Dimock has shown that any isometry $\theta: {\bf M}_1 \to {\bf M}_2$
induces a $C^*$-algebraic isomorphism $\a_{\theta}: \fA[K_{{\bf
    M}_1},\k_{{\bf M}_1}] \to \fA[K_{{\bf M}_2},\k_{{\bf M}_2}]$ with
the property that
\begin{equation}
\label{ex.i}
 \a_{\theta}(W_{{\bf M}_1}([f]_{{\bf M}_1})) = W_{{\bf
     M}_2}([\theta^*f]_{{\bf M}_2})\,, \quad f \in
 C_0^{\infty}(M_1,\bR)\,,
\end{equation}
where $\theta^*f = f \lcrc \theta^{-1}$. If ${\bf M}_1 \subset {\bf
  M}_1'$ and ${\bf M}_2 \subset {\bf M}_2'$ are globally hyperbolic
sub-spacetimes of a pair of globally hyperbolic spacetimes ${\bf M}_1'$
and ${\bf M}_2'$, then $W_{{\bf M}_j} = W_{{\bf M}_j'}\rest K_{{\bf
    M}_j}$ ($j=1,2$) holds up to $C^*$-isomorphisms as a consequencs
of the uniqueness of the causal propagators, thus there is always a
$C^*$-algebraic Weyl-algebra isomorphism  covering a
local isomorphism between members of $\cG$. Furthermore, Dimock
has also shown in \cite{Dim.KG} that, upon denoting by $\fA_{\bf
  M}(O)$ the $C^*$-subalgebra of $\fA[K_{\bf M},\k_{\bf M}]$ generated
by all $W_{\bf M}([f]_{\bf M})$, ${\rm supp}\,f \subset O$, there
holds
\begin{equation}
\label{ex.ii}
 O_1 \ind O_2 \imply \fA_{\bf M}(O_1) \subset \fA_{\bf M}(O_2)
\end{equation}
for all $O_1,O_2 \subset M$.

Now let $\o_{\bf M}$ be an arbitrary quasifree Hadamard state on
$\fA[K_{\bf M},\k_{\bf M}]$. Such a state is determined by its
two-point correlation function which here is required to be of
``Hadamard form''. The Hadamard form specifies the singular
short-distance behaviour in a particular way, see \cite{Ful,WaldII}
and references cited therein for discussion. Equivalently, the
Hadamard form of a two-point function can be characterized by a
certain form of its wavefront set (see \cite{Rad1,SaV2} for
details). It has been shown in \cite{FNW} that there exists an
abundance of Hadamard states on $\fA[K_{\bf M},\k_{\bf M}]$. To such a
quasifree Hadamard state $\o_{\bf M}$ there corresponds its
GNS-Hilbertspace representation $(\pi_{\bf M},\cH_{\bf M},\O_{\bf
  M})$, cf.\ e.g.\ \cite{BR1}. In that representation, we define the
local von Neumann algebras 
$$\cF_{\bf M}(O) = \pi_{\bf M}(\fA_{\bf M}(O))'' $$
for each $O \in {\rm orc}(M)$.
Then \eqref{ex.ii} clearly implies the existence of a causal dynamical
law
$$  O_1 \ind O_2 \imply \cF_{\bf M}(O_1) \subset \cF_{\bf M}(O_2)\,.$$
 
A vector $\chi \in \cH_{\bf M}$ is
defined to be in $\cD_{\bf M}$ if for each choice of $\vec{x} =
(x_1,\ldots,x_n) \in (K_{\bf M})^n$ the map
$$ \vec{t} \mapsto \pi_{\bf M}(W_{\bf M}(t_1x_1))\cdots \pi_{\bf
  M}(W_{\bf M}(t_nx_n))\chi\,, \quad \vec{t} = (t_1,\ldots,t_n) \in
\bR^n\,,
 $$
is $C^{\infty}$. One can show that $\cD_{\bf M}$ is a dense domain in
$\cH_{\bf M}$ (cf.\ \cite{BR2}). One can define for each $f \in
C_0^{\infty}(M,\bR)$ the quantum field operator $\Fi_{\bf M}(f)$ by
$$ \Fi_{\bf M}(f)\chi = \left.-i\frac{d}{dt}\right|_{t = 0}\pi_{\bf
  M}(W_{\bf M}(t[f]_{\bf M}))\chi\,, \quad \chi \in \cD_{\bf M}\,.
 $$
One can also show that $\cD_{\bf M}$ is left invariant under the
action of $\Fi_{\bf M}(f)$ and that $\Fi_{\bf M}(f)$ is essentially
self-adjoint \cite{BR2}. It is also obvious that $\Fi_{\bf M}(f)$ is
affiliated to $\cF_{\bf M}(O)$ as soon as ${\rm supp}\,f \subset O$.

Moreover, the results of \cite{Ver1} show that the $C^*$-algebraic
isomorphism $\alpha_{\theta}$ in \eqref{ex.i} can be extended, in
representations induced by quasifree Hadamard states, to von Neumann
algebraic isomorphisms in the following way. Suppose that between ${\bf
  M}_1$ and ${\bf M}_2$ in $\cG$ there is a local isomorphism $\theta$,
and let $N_i \subset \ell_{\rm
  ini}(\theta)$ be a relatively compact subset. Then, writing $N_f =
\theta(N_i)$, the Weyl-algebra
isomorphism $\alpha_{\theta}$ in \eqref{ex.i} extends to an
isomorphism $\alpha_{\theta,N_i} : \cF_{{\bf M}_1}(N_i) \to \cF_{{\bf
    M}_2}(N_f)$ between von Neumann algebras. Consequently, there
holds the covariance property
$$ \a_{\theta,N_i}(\cF_{{\bf M}_1}(O_i)) = \cF_{{\bf
    M}_2}(\theta(O_i))\,, \quad O_i \in {\rm orc}(N_i)\,.$$
Finally, if ${\bf M}_0$ is Minkowski spacetime, we take $\o_{{\bf
    M}_0}$ to be the vacuum state which is known to be a quasifree
Hadamard state. In conclusion, the just constructed family
$\{\FFi_{\bf M}\}_{{\bf M} \in \cG}$ of Klein-Gordon quantum fields
for each ${\bf M} \in \cG$ satisfies all the assumptions required for
a generally covariant quantum field theory over $\cG$.
\\[10pt]
{\it 2. The Proca field. } The Proca field is a co-vector field, i.e.\
of tensorial type, corresponding to the $D^{(1,1)}$ irreducible
representation of SL$(2,\bC)$. For each globally hyperbolic spacetime
${\bf M} = 
(M,g) \in \cG$ (where again we suppress the spin-structure in our
notation since it is presently not relevant), we denote by $d$ the
exterior derivative of differential forms, by $*$ the Hodge-star
operator corresponding to the metric $g$, and define the
co-differential $\d = *d*$. Then the Proca equation reads, for
$\varphi \in \G_0(T^*M)$,
$$ (\d d + m^2)\varphi = 0\,,$$
where $m > 0$ is a constant independent of ${\bf M}$. (Note that $\d
d$ depends on the metric $g$.) A $C^*$-algebraic quantization has
recently been given by Furlani \cite{Fur} (cf.\ also \cite{Stro},
whose notation we follow here). To this end one constructs advanced
and retarded fundamental solutions $F_{\bf M}^{\pm} : \G_0(T^*M) \to
\G(T^*M)$ uniquely determined by
$$ F_{\bf M}^{\pm}(\d d + m^2)f = f = (\d d + m^2)F^{\pm}_{\bf M}f\,,
\ \ {\rm supp}\,F^{\pm}_{\bf M}f \subset J^{\pm}({\rm supp}\,f)\,,
\quad f \in \G_0(T^*M)\,.$$
As in the case of the scalar Klein-Gordon field, one defines the
(causal) propagator $F_{\bf M} = F^+_{\bf M} - F^-_{\bf M}$ and a
symplectic space $(K_{\bf M},\k_{\bf M})$ where
$$ \k_{\bf M}([f],[h]) = \int_M f \wedge *F_{\bf M}h\,, \quad [f],[h]
\in K_{\bf M}\,,$$
on $K_{\bf M}= \G_0(T^*M)/{\rm ker}\,F_{\bf M}$ and $f \mapsto [f] =
[f]_{\bf M}$ is the quotient map.

From here onwards, all the arguments leading to the construction of a
generally covariant theory $\{\FFi_{\bf M}\}_{{\bf M}\in \cG}$ can be
taken over almost literally, except for obvious modifications,
from the previous case of the scalar
Klein-Gordon field to the present case of the Proca field. There are
some provisions which should nevertheless be recorded: Firstly, the
existence of Hadamard states for the Proca field has not been
demonstrated. However, as mentioned towards the end of Sec.\ 5.1 in
\cite{SaV2}, the existence of Hadamard states could be established by
using the existence of a ground state for the Proca field on
ultrastatic spacetimes \cite{Fur.ul} in combination with results in
\cite{SaV1} and \cite{FNW} to prove that there exists a large set of
quasifree Hadamard states for the Proca field. Secondly, the arguments
given in \cite{Ver1} showing that the $C^*$-algebraic isomorphism
\eqref{ex.i} can be extended to a von Neumann algebraic isomorphism in
the above said way apply to the case of the free scalar
Klein-Gordon field. But those arguments can obviously be generalized to
apply to a far more general class of free fields, including the Proca
field. Thus, one may conclude that also the Proca field gives rise to
a generally covariant quantum field theory $\{\FFi_{\bf M}\}_{{\bf M}\in
\cG}$. 
\\[10pt]
{\it 3. The Dirac field. } Our last example is the Dirac field, which
is a spinorial field of spin $1/2$. We consider it in a Majorana
representation; our presentation follows \cite{Dim.D} to large extent,
with some alterations specific to Majorana representations, see
\cite{SaV2} for details. The Majorana representation corresponds to
the real linear irreducible representation $D^{(1,0)}
\oplus D^{(0,1)}$ of SL$(2,\bC)$. This  Majorana-Dirac 
representation will be denoted by
$\rho$. Its representation space is $V_{\rho} = \bC^4$.

Let ${\bf M} = (M,g, S(M,g), \psi) \in \cG$ be a globally hyperbolic
spacetime with spin-structure. The vector bundle $\cV = S(M,g) \ltimes_{\rho}
\bC^4$ associated with $S(M,g)$ and the representation $\rho$ will be
denoted by $D_{\rho}M$; its sections are called spinors, or spinor
fields.
The metric-induced connection $\nabla$ on $TM$ lifts to a connection
on the frame bundle $F(M,g)$ which in turn lifts to a connection on
$S(M,g)$, and this induces also a connection on $D_{\rho}M$. The
corresponding covariant derivative operator will be denoted by
$\fatnabla$. One can then introduce the spinor-tensor
$\boldsymbol{\gamma} \in \G(T^*M \otimes D_{\rho}M \otimes
D^*_{\rho}M)$ by requiring that its components
$\boldsymbol{\g}_{a}{}^A{}_B$ in (appropriate, dual) local frames
are equal to the matrix elements $(\g_{a})^A{}_B$ of the
gamma-matrices in the Majorana-representation. This is a set of four $4
\times 4$ matrices $\g_0,\g_1,\g_2,\g_3$ obeying the relations
$$
     \g_{a}\g_{b} + \g_{b}\g_{a} = 2\eta_{ab}\,, \quad
\g_0^* = \g_0\,, \quad \g^*_k = -\g_k\ (k = 1,2,3)\,, \quad
\overline{\g_{a}} = \g_{a}\,.
 $$
Here, $\g_{a}^*$ means the Hermitean conjugate of $\g_{a}$ and
$\overline{\g_{a}}$ is the transpose of $\g_{a}^*$, and
$(\eta_{ab}) = {\rm diag}(1,-1,-1,-1)$ is the Minkowskian metric.
Then the Dirac-operator $\dirop$ is defined by setting in frame
components, for any local section $f = f^AE_A \in \G_0(D_{\rho}M)$,
$$ (\dirop f)^A = \eta^{a
  b}{\boldsymbol{\g}}_{a}{}^A{}_B(\fatnabla_{b} f)^B\,.
  $$
(At this point, we refer to \cite{Dim.D,SaV2} for details.) There is a
charge conjugation $C$ which operates by complex conjugation of the
frame-components in any frame, i.e.\ $(Cu)^A = \overline{u^A}$ for the
components of $u \in D_{\rho}M$. There is also the Dirac adjoint $u
\mapsto u^+$
mapping $D_{\rho}M$ anti-linearly and base-point preserving onto its
dual bundle $D^*_{\rho}M$; in dual frame components it is defined as
$(u^+)_B = \overline{u^A}\g_{0\,AB}$. 

The Dirac-equation on ${\bf M}$ is the differential equation
$$ (\dirop + im)\varphi = 0$$
for $\varphi \in \G(D_{\rho}M)$ where $m \ge 0$ is a constant,
independent of ${\bf M}$. As in the cases considered before,
there are uniquely determined advanced and retarded fundamental
solutions $S^{\pm}_{\bf M} : \G_0(D_{\rho}M) \to \G(D_{\rho}M)$ distinguished
by the properties
$$ S^{\pm}_{\bf M}(\dirop + im)f = f = (\dirop + im)S^{\pm}_{\bf
  M}f\,,\ \  {\rm supp}\,S^{\pm}_{\bf M}f \subset J^{\pm}({\rm
  supp}\,f)\,, \quad f \in \G_0(D_{\rho}M)\,.$$
Hence one obtains a distinguished causal propagator $S_{\bf M} =
S^+_{\bf M} - S^-_{\bf M}$.
It gives rise to a pre-Hilbertspace $(H_{\bf M},s_{\bf M})$ where
$H_{\bf M} = \G_0(D_{\rho}M)/{\rm ker}\,S_{\bf M}$ with scalar product
$$ s_{\bf M}([f],[h]) = \int_M d\eta\,(Sf)^+(h)\,, \quad [f],[h] \in
H_{\bf M}\,,$$
where we have denoted the metric-induced measure on $M$ by $d\eta$ and
by $f \mapsto [f] = [f]_{\bf M}$ the quotient map. The charge
conjugation $C$ can be shown to induce a conjugation on $(H_{\bf
  M},s_{\bf M})$ which will be denoted by the same symbol. We shall
also notationally identify $H_{\bf M}$ with its completion to a
Hilbertspace.

To the Hilbertspace $(H_{\bf M},s_{\bf M})$ with complex conjugation
$C$ there corresponds (uniquely, up to $C^*$-algebraic equivalence)     
the self-dual CAR-algebra $\fB[H_{\bf M},s_{\bf M},C]$ (cf.\
\cite{Ara}) which is a $C^*$-algebra generated by elements $B_{\bf
  M}(v)$ depending linearly on $v \in H_{\bf M}$ and
 fulfilling the canonical anti-commutation
relations
$$ B_{\bf M}(v)^*B_{\bf M}(w) + B_{\bf M}(w)B_{\bf M}(v)^* = s_{\bf
  M}(v,w)\,, \ \ B_{\bf M}(v)^* = B_{\bf M}(Cv)\,, \quad v,w \in
H_{\bf M}\,.$$
In \cite{Dim.D}, Dimock has proven that each (global) isomorphism $\TTh =
(\Theta,\vartheta)$ between members ${\bf M}_1$ and ${\bf M}_2$ in
$\cG$ induces a $C^*$-algebraic isomorphism $\a_{\TTh} : \fB[H_{{\bf
    M}_1},s_{{\bf M}_1},C] \to \fB[H_{{\bf M}_2},s_{{\bf M}_2},C]$
satisfying
\begin{equation}
\label{ex.iii}
 \a_{\TTh}(B_{{\bf M}_1}([f]_{{\bf M}_1})) = B_{{\bf
    M}_2}([\check{\Theta}^{\star}f]_{{\bf M}_2})\,, \quad f \in
\G_0(D_{\rho}M_1)\,,
\end{equation}
where $\check{\Theta}^{\star}f = \check{\Theta}\lcrc f \lcrc
\vartheta^{-1}$, $\check{\Theta}$ being the map $D_{\rho}M_1 \to
D_{\rho}M_2$ induced by $\Theta$. As in the cases discussed before,
this statement has a local version to the effect that for each local
isomorphism between members of $\cG$ there is a $C^*$-algebraic
isomorphism between the corresponding CAR-algebras covering it.

Moreover it was shown in \cite{V.diss} that strong Einstein causality,
\begin{equation} \label{ex.iv}
O_1 \ind O_2 \imply \fB_{\bf M}(O_2) \subset \fB_{\bf M}(O_2)\,, 
\end{equation}
holds for the local $C^*$-subalgebras $\fB_{\bf M}(O)$ of $\fB[H_{\bf
  M},s_{\bf M},C]$ which are generated by all $B_{\bf M}([f]_{\bf M})$ with 
${\rm supp}\,f \subset O$.       

Now let $\o_{\bf M}$ be any quasifree Hadamard state on $\fB[H_{\bf
  M},s_{\bf M},C]$, and $(\pi_{\bf M},\cH_{\bf M},\O_{\bf M})$ the
corresponding GNS-representation, then the local von Neumann algebras
will be defined via
$$ \cF_{\bf M}(O) = \pi_{\bf M}(\fB_{\bf M}(O))''\,, \quad O \in {\rm
  orc}(M)\,,$$ 
whereas the field operators are now given as
$$ \Fi_{\bf M}(f) = \pi_{\bf M}(B_{\bf M}([f]_{\bf M}))\,, \quad f \in
\G_0(D_{\rho}M)\,.$$
Owing to the canonical anti-commutation relations, these field operators are
bounded, and one may take their domain $\cD_{\bf M}$ to be equal to
$\cH_{\bf M}$. The existence of a causal dynamical law at the level of
the local von Neumann algebras is then granted by \eqref{ex.iv}.

It is to be expected that the arguments of \cite{Ver1} showing that
the $C^*$-algebraic Weyl-algebra isomorphisms \eqref{ex.i} (when
appropriately localized, see above) extend to von Neumann algebraic
isomorphisms for the case of the scalar Klein-Gordon field have
generalizations allowing to conclude that the $C^*$-algebraic
CAR-algebra isomorphisms \eqref{ex.iii} extend, in a similar manner,
to von Neumann algebraic isomorphisms, so that general covariance is
fulfilled. Another provision is that, as in the case of the Proca field,
the existence of quasifree Hadamard states for the Dirac field has as
yet not been demostrated. However, the same comment as given above for
the case of the Proca field applies here. Anticipating therefore that
these provisions are lifted, the just constructed family $\{\FFi_{\bf
  M}\}_{{\bf M}\in \cG}$ of Dirac quantum fields for each ${\bf M} \in
\cG$ yields another example of a generally covariant quantum field
theory over $\cG$ upon choosing $\o_{{\bf M}_0}$ as the vacuum state
(being quasifree and Hadamard) on Minkowski spacetime ${\bf M}_0$.

${}$\\[30pt]
\noindent
{\Large \bf Appendix A}
\renewcommand{\theequation}{\Alph{section}.\arabic{equation}}
\setcounter{section}{1}
\setcounter{equation}{0}
\\[24pt]
{\it Proof of Lemma 2.1. } Let two causally separated points $p_1$ and
$p_2$ be given; hence we may form the manifold $M^{\vee} =
M\backslash(J^+(p_1) \cup J^+(p_2))$. Then
$(M^{\vee},g\rest M^{\vee})$ is again a globally hyperbolic
spacetime. This globally hyperbolic spacetime may be smoothly foliated
into Cauchy-surfaces and thus one can move Cauchy-surfaces for
$(M^{\vee},g\rest M^{\vee})$ arbitrarily close to $p_1$ and
$p_2$. We will use this property in order to construct a
Cauchy-surface $\Sigma$ in $(M,g)$ having the following properties:
\begin{itemize}
\item[(i)] $\Sigma \subset M^{\vee}$
\item[(ii)] There is an open, simply connected neighbourhood $W
  \subset \Sigma$ which is contained in a coordinate chart (for
  $\Sigma$), and it holds that $J^-(p_j) \cap \Sigma \subset W$ ($j=1,2$).
\end{itemize}
To this end, let $F:\bR \times \Sigma_0 \to M$ be a $C^{\infty}$-foliation of
$(M,g)$ in Cauchy-surfaces. If $C$ is any Cauchy-surface in
$(M,g)$, then there is a diffeomorphism $\Psi_{C} : \Sigma_0 \to C$
which is defined by assigning to $x \in \Sigma_0$ the point $q_x \in
C$ so that $F(t_x,x) = q_x$ for some (uniquely determined) $t_x
\in \bR$. Now let $(t_j,x_j) \in \bR \times \Sigma_0$ be such that
$F(t_j,x_j) = p_j$, $j =1,2$. Then there is clearly a pair
$S_1,S_2$ of open neighbourhoods of $x_1,x_2$, respectively, in
$\Sigma_0$ lying in a simply connected chart domain $W_0$ (of
$\Sigma_0$), cf.\ \cite{Dieu3}, Prop.\ 16.26.9. Thus, whenever $C$ is a
Cauchy-surface in $(M,g)$, then the sets $\Psi_C(S_1)$ and
$\Psi_C(S_2)$ are contained in the simply connected chart domain
$\Psi_C(W_0)$ of $C$. On the other hand, $\Psi_C(S_j)$ is the
intersection of $C$ with the `tube' $T_j = \bigcup \{F(t,x):
t\in\bR,\ x \in S_j\}$. It is now fairly easy to see that, if
$B_j$ denotes the unit ball in $T_{p_j}M$ with respect to arbitrarily
given coordinates, then the sets $V_j(\tau)=\{\exp_{p_j}(v): v\in
\tau \cdot B_j,\ v\ \mbox{\rm past-directed\ and\ causal}\}$ of segments of
`causal rays' emanating to the past from $p_j$ 
will be contained in $T_j$ if $\tau > 0$ is small enough. Choosing
such a $\tau$ and using that
$(M^{\vee},g\rest M^{\vee})$ is globally hyperbolic, one can thus
find a Cauchy-surface $\Sigma$ in $(M^{\vee},g\rest M^{\vee})$ with
$(V_j(\tau)\backslash V_j(\tau/2)) \subset {\rm int}\,J^-(\Sigma)$; this 
implies that the intersection of $J^-(p_j)$ with $\Sigma$ is contained
in $T_j \cap \Sigma = \Psi_{\Sigma}(S_j)$, and since $\Sigma$ is
also a Cauchy-surface for $(M,g)$, one realizes that it has the desired
properties (i) and (ii) upon choosing $W = \Psi_{\Sigma}(W_0)$.  

In a next step we note that, since the sets $J^-(p_j) \cap \Sigma$ are
closed and contained in the open set $W$, also the closures of
sufficiently small open neighbourhoods of these sets are contained in
$W$. Thus we can choose two sufficiently small sets $U_j = {\rm
  int}(J^-(p_1^+)\cap J^+(p_j^-))$ where $p_j^{\pm} \in {\rm
  int}\,J^{\pm}(p_j)$, i.e.\ they are `double cones' surrounding the points
$p_j$, with $\overline{J^-(U_j)} \cap \Sigma \subset W$.
[Note that in Figure 1 we have represented the sets $U_j$ as truncated
double cones since this turned out be be easier graphically.]
 Obviously
one may choose the $U_j$ so that they are contained in $N_+ =
{\rm int}\,J^+(\Sigma)$. Moreover, $J^-(U_j)\cap \Sigma$ will be
contained in an open, simply connected subset $W_1$ of $\Sigma$ with
$\overline{W_1} \subset W$. Then ${\rm int}\,D^+(W_1)$ is a simply
connected neighbourhood of $\overline{U_1}$ and $\overline{U_2}$, 
 and is globally
hyperbolic when endowed with the metric $g$. Since
$(N_+,g\rest N_+)$ is a globally hyperbolic spacetime, one can
choose a Cauchy-surface 
$\Sigma_+$ in $(N_+,g\rest N_+)$ `sufficiently close to $\Sigma$' 
so that the set
$$ G = {\rm int}\,D^+(W_1) \cap {\rm int}\,J^+(\Sigma_+)\subset N_+ $$
is still an open, simply connected neighbourhood of $\overline{U_1}$ and
$\overline{U_2}$ which is globally hyperbolic when supplied with $g$
as metric.    

The remaining part of the argument proceeds in a similar way as the
proof of Appendix C in \cite{FNW}. We can cover $\Sigma$ with a system
$\{X_{\a}\}_{\a}$ of coordinate patches, choosing one of them, say
$X_1$, to have the property
\begin{equation}
\label{Xincl}
\overline{W}_1 \subset X_1\,, \quad \overline{X}_1 \subset
W\,.
\end{equation}
Using Gaussian normal coordinates for $\Sigma$, one may introduce
coordinate patches $(-\varepsilon_{\a},\varepsilon_{\a})\times X_a$
covering a neighbourhood $N_0$ of $\Sigma$, on each of which the
metric $g$ assumes the form
$$   dt^2 - g_{ij}(t,x)dx^idx^j
 $$
where $t \in (-\varepsilon_{\a},\varepsilon_{\a})$ and
$x=(x^i)_{i=1}^3$ are coordinates on $X_{\a}$; $(g_{ij}(t,x))$ are the
coordinates of the $3$-dim.\ Riemannian metric induced by the metric
$g$ on the slices of constant $t$. Here, the coordinatization is
assumed to be such that $(t,x)$ represents a point in $N_+$ for $t >
0$ and a point in $N_- = {\rm int}\,J^-(\Sigma)$ for $t
<0$. Moreover, $N_0$ may be chosen so that it is, with $g\rest
N_0$ as metric, a globally hyperbolic sub-spacetime of $(M,g)$,
and assuming now that $N_0$ has been chosen in that way, also $N_0
\cap N_-$ is a globally hyperbolic sub-spacetime with the appropriate
restriction of $g$ as metric. After a moment of reflection one can see
that this implies the existence of a Cauchy-surface $\Sigma_1$ in $N_0
\cap N_-$ so that
$$ J^-(\overline{W}_1) \cap J^+(\Sigma_1) \subset (-\varepsilon_1,0)
\times X_1$$
by `moving $\Sigma_1$ sufficiently close to $\Sigma$'. Upon moving
$\Sigma_1$, if necessary, `still closer' to $\Sigma$, it is also
possible to ensure that the parts of $J^-(\overline{U}_1)$ and
$J^-(\overline{U}_2)$ lying in $J^+(\Sigma_1)$ are causally
separated. With $\Sigma_1$ chosen in that manner, one can now pick
some pair of small neighbourhoods $\widetilde{U}_j$ lying relatively
compact in ${\rm int}(J^+(\Sigma_1)\cap J^-(U_j))$ ($j=1,2$). We may
then also select another Cauchy-surface $\Sigma_2$ in $N_0 \cap
N_1$, with
$$ {\rm cl}\,\widetilde{U}_j \subset {\rm int}\,J^-(\Sigma_2)\,,
\quad \Sigma_2 \subset {\rm int}\,J^+(\Sigma_1)\,.$$

In the next step, we endow $\Sigma$ with a complete Riemannian metric
$\gamma$, which we prescribe to be a flat Euclidean metric on $X_1$
(which is possible because of \eqref{Xincl} in view of the fact that
$W$ is a coordinate patch). We shall, furthermore, choose $\g$ so
that the flat Lorentzian metric $\eta$ on $(-\varepsilon_1,0) \times
X_1$ given by
$$ \eta = dt^2 - \gamma_{ij}dx^idx^j $$
has for $(t,x) \in (-\varepsilon_1,0) \times \cX_1$ the property that
each causal curve for $\eta$ is also a causal curve for $g$, i.e.\
$J_{\eta}(q) \subset J_g(q)$ on $(-\varepsilon_1,0) \times X_1$. This may
always be realized by rescaling $\gamma$ by a constant factor.

Now define $\widetilde{M} = {\rm int}\,J^+(\Sigma_1)$. Let $f
\in C^{\infty}(\widetilde{M},\bR_+)$ have the following properties:
$0 \le f \le 1$, $f \equiv 0$ on $J^+(\Sigma)$, $f \equiv 1$ on
$J^-(\Sigma_2)$. Then define a metric $\widetilde{g}$ on $N_0 \cap
\widetilde{M}$ by setting its coordinate expression to be equal to
$$ b(t,x)dt^2 - \left( f(t,x) \g_{ij} +
  (1-f(t,x))g_{ij}(t,x)\right)dx^idx^j $$
on each coordinate patch $(-\varepsilon_{\a},\varepsilon_{\a})\times
X_{\a}$. Here, $b$ is a smooth function on $N_0
\cap \widetilde{M}$ with $0 < b \le 1$ and sufficiently 
small so that, with the new
metric $\widetilde{g}$, $N_0$ is globally hyperbolic; from the
properties of $\g$ mentioned before it is obvious that one can choose
such a $b$ so that $b \equiv 1$ on $N_+$ and $b \equiv 1$ on the set
$$ Y = {\rm int}\left(\widetilde{M} \cap J^-(\Sigma_2) \cap
  (-\varepsilon_1,0) \times X_1\right)\,.$$
With this choice of $b$, it is moreover clear that $\widetilde{g}$
coincides on $N_+$ with the metric $g$, and so $\widetilde{g}$ may
be extended from $N_0\cap \widetilde{M}$ to all of
$\widetilde{M}$ by defining $\widetilde{g}$ as $g$ on
$N_+$. Moreover, $\widetilde{g}$ is a flat Lorentzian metric on
$Y$, and viewing $U_j$, $j =1,2$, canonically as subsets of
$\widetilde{M}$, the previous contructions entail that there are two
globally hyperbolic sub-spacetimes $\widehat{U}$ (with metric
$\widetilde{g}$) which are relatively compact in $Y$, and have the
property that $\widehat{U}_j \dni U_j$ with respect to the metric
$\widetilde{g}$.

Finally, one can make $Y$ slightly smaller in order to obtain a
globally hyperbolic sub-spacetime $\widehat{G}$ of
$(\widetilde{M},\widetilde{g})$ which is simply connected and still
contains $\widetilde{U}_j$ and $\widehat{U}_j$ (if necessary, by
making the $\widehat{U}_j$ slightly smaller as well); and $\widetilde{g}$ is
flat on $\widehat{G}$. Therefore we have now constructed the
required $(\widetilde{M},\widetilde{g})$ and the subsets $U_j$,
$\widetilde{U}_j$, $\widehat{U}_j$ ($j=1,2$) and $G$,
$\widehat{G}$ with the properties claimed in Lemma 1.
\hfill $\Box$    
\\[30pt]
{\large \bf Appendix B}
\setcounter{section}{2}
\setcounter{equation}{0}
\\[24pt]
In this appendix we collect the assumptions about a quantum field
theory $\FFi_{{\bf M}_0}$ on Minkowski spacetime equipped with its
standard spin structure, and quote the spin-statistics theorem for
this setting. The assumptions are those given in the book
by Streater and Wightman \cite{StrWi}, except that in formulating the
Bose-Fermi alternative (normal commutation relations), we will posit
that Bosonic commutation relations hold in the strong sense, similarly
as in the statement of Thm.\ 4.1. See below for details.

To begin with, write $(M_0,\eta) = (\bR^4,{\rm diag}(+,-,-,-))$ for
Minkowski spacetime. A Lorentzian coordinate frame $(e_0,\ldots,e_3)$
has been chosen by which $M_0$ is identified with $\bR^4$, and which
also serves to fix orientation and time-orientation. The framebundle
$F(M_0,\eta)$ is isomorphic to $\bR^4 \semidirprod \cL^{\uparrow}_+$,
and for each $x \in \bR^4$, $(x,(e_0,\ldots,e_3))$ represents an
element in $F(M_0,\eta)$. Then the spin-bundle $S(M_0,\eta)$ is
isomorphic to $\bR^4 \semidirprod {\rm SL}(2,\bC)$, and one obtains a
spin-structure $\psi_0: S(M_0,\eta) \to F(M_0,\eta)$ by assigning to
$(x,{\bf s}) \in S(M_0,\eta)$ the element $\psi_0(x,{\bf s}) =
(x,(e_0({\bf s}),\ldots,e_3({\bf s})))$ in $F(M_0,\eta)$ with
$$ e_b({\bf s}) = e_a \L^a{}_b({\bf s})$$
where ${\rm SL}(2,\bC) \owns {\bf s} \mapsto \L({\bf s}) \in
\cL^{\uparrow}_+$ is the covering projection. Explicitly, the matrix
components of $\L({\bf s})$ are given by
$$ \L_{ab}({\bf s}) = \frac{1}{2}{\rm Tr}({\bf s}^*\sigma_a{\bf
  s}\sigma_b)$$ 
where $\sigma_0,\ldots,\sigma_3$ are the Pauli-matrices.

Now let $\rho$ denote any of the complex linear irreducible
representations $D^{(k,l)}$, or of the real linear irreducible
representations $D^{(k,l)}\oplus D^{(l,k)}$ $(k \ne l)$, where $k,l
\in \bN_0$. The corresponding representation space will be denoted by
$V_{\rho}$. 
 Then we
require that the quantum field theory $\FFi_{{\bf M}_0} = (\Fi_{{\bf
    M}_0},\cD_{{\bf M}_0},\cH_{{\bf M}_0})$ has the following
properties (where in the following, we abbreviate $(\Fi_{{\bf
    M}_0},\cD_{{\bf M}_0},\cH_{{\bf M}_0})$ by $(\Fi_0,\cD_0,\cH_0)$):
\begin{itemize}
\item[1.)] $\cH_0$ is a Hilbertspace and $\cD_0 \subset \cH_0$ is a
  dense linear subspace.
\item[2.)] $\Fi_0$
  is a linear map taking elements $f$ in $\Srv$ to closable operators
  $\Fi_0(f)$ all having the common, dense and invariant domain $\cD_0$. Here,
  $\Srv$ is the set of Schwartz-functions on $\bR^4$ taking values in
  the finite-dimensional representation space $V_{\rho}$.
\footnote{In the case of flat Minkowski-spacetime, $S(M_0,\eta) =
  \bR^4 \times {\rm SL(2,\bC)}$ and one can canonically identify
  $\cV_{\rho}$ with $\bR^4 \times V_{\rho}$ and $\check{\rho}$ with
  ${\rm id} \times \rho$.}
\item[3.)] For each pair of vectors $\chi,\chi' \in \cD_0$, the map
$$ \Srv \owns f \mapsto (\chi,\Fi_0(f)\chi') $$
is continuous, hence an element in ${\cS}'(\bR^4,V_{\rho})$.
\item[4.)] There is a strongly continuous representation 
  $$ \widetilde{\cP}^{\uparrow}_+ \owns (a,{\bf s}) \mapsto U(a,{\bf
    s})$$
of $\widetilde{\cP}^{\uparrow}_+ = \bR^4 \rtimes {\rm SL}(2,\bC)$
(the covering group of the proper orthochronous Poincar\'e group) by
unitary operators on $\cH_0$; $\cD_0$ is left invariant under the
action of the $U(a,{\bf s})$.
\item[5.)] The spectrum of the translation-subgroup $a \mapsto U(a,1)$
  is contained in the closed forward lightcone $\overline{V}_+$, i.e.\
  the relativistic spectrum condition holds. Moreover, there is an up
  to a phase unique unit vector $\O \in \cH_0$, the vacuum vector,
  fulfilling $U(a,{\bf s})\O = \O$ for all $(a,{\bf s}) \in
  \widetilde{\cP}^{\uparrow}_+$. This vector is assumed to be
  contained in $\cD_0$ and to be cyclic
  for the algebra generated by the field operators in the sense that
  $\cD_0$ coincides with the vector space spanned by $\O$ and all
  vectors of the form $F_1\cdots F_n\O$, $n \in \bN$,
   $F_j \in \{\Fi_0(f_j),\Fi_0(f_j)^*\}$, $f_1,\ldots,f_n \in \Srv$.
\item[6.)] The quantum field possesses the covariance property
 $$U(a,{\bf s})\Fi_0(f)U(a,{\bf s})^{-1} =
 \Fi_0(\rho_a{}^{\star}({\bf s})f) \,,$$
where
$$ \rho_a{}^{\star}({\bf s})f(y) = \rho({\bf s})(f(\L({\bf
  s})^{-1}(y -a)))$$
for all $a \in \bR^4$, ${\bf s} \in {\rm SL}(2,\bC)$, $f \in \Srv$.
\item[ 7.)] Spacelike clustering holds on the vacuum, i.e.\ if $a$ is
  any non-zero spacelike vector, then one has
\begin{eqnarray*}
& & (\O,F_1\cdots F_kU(ta,1)F_{k+1} \cdots
 F_n\Omega)\\
& & \quad \underset{t \to \infty}{\longrightarrow}
 (\O,F_1 \cdots
 F_k\O)(\O,F_{k+1}\cdots F_n\O) 
\end{eqnarray*}
for all $F_j \in \{\Fi_0(f_j),\Fi_0(f_j)^*\}$, with $f_1,\ldots,f_n
\in \Srv$, $n \in \bN$. 
\item[8.)] Finally, the Bose-Fermi alternative is required to hold in
  the following form. The quantum field fulfills either
\\[6pt]
{\it Bosonic commutation relations}:\\[4pt]
Given any pair of causally separated subsets $O_1,O_2 \in {\rm
  orc}(\bR^4)$, then it holds that 
$$ \cF_0(O_1) \subset \cF_0(O_2)'\,,$$
or
\\[6pt]
{\it Fermionic commutation relations}:\\[4pt]
Given any pair of $f_1,f_2 \in \Srv$ with spacelike separated
supports, then it holds that
$$ \Fi_0(f_1)\Fi_0(f_2) + \Fi_0(f_2)\Fi_0(f_1) = 0 \ \ {\rm and}\ \ 
\Fi_0(f_1)\Fi_0(f_2)^* + \Fi_0(f_2)^*\Fi_0(f_1) = 0\,.$$
\end{itemize}
In formulating the statement of Bosonic commutation relations (or {\it
  locality}, as it is also called), $\cF_0(O)$ denotes the von Neumann
algebra generated via the polar decomposition of the closed field
operators $\overline{\Fi_0(f)}$ with ${\rm supp}\,f \subset O$ as
described in assumption (a) of Sec.\ 4. The above statement of Bosonic
commutation relations is thus equivalent to saying that the field
operators $\Phi_0(f_1)$ and $\Phi_0(f_2)$ commute strongly for
spacelike separated supports of $f_1$ and $f_2$; here we say that a
pair of closable operators $X_j$ $(j= 1,2)$ commutes strongly if
$J_1$ and ${\rm e}^{is|X_1|}$ commute with $J_2$ and ${\rm
  e}^{it|X_2|}$, $s,t \in \bR$, where $\overline{X}_j = J_j|X_j|$
denotes polar decomposition. Clearly, the property of field operators
to commute strongly at spacelike separation implies their spacelike
commutativity in the ordinary sense,
$$
 \Fi_0(f_1)\Fi_0(f_2) - \Fi_0(f_2)\Fi_0(f_1) = 0\ \ {\rm and}\ \
 \Fi_0(f_1)\Fi_0(f_2)^* - \Fi_0(f_2)^*\Fi_0(f_1) = 0
  $$
whenever the supports of $f_1$ and $f_2$ are spacelike separated, but
without further information one can in general not conclude that this
last relation also implies spacelike commutativity of the field
operators in the strong sense as usually the field operators will be
unbounded. The question as to when this conclusion may nevertheless be
drawn for field operators in quantum field theory is a longstanding
one; however, several criteria are known. We refer the reader to
\cite{BorYng, DriSumWi} for further 
discussion and references. Suffice it to say here that ordinary
spacelike commutativity is expected to imply strong spacelike
commutativity of field operators in the case of physically relevant
theories.

We also mention that in Def.\ 4.1 the quantum field $\Fi_0 =\Fi_{{\bf M}_0}$
has only been assumed to be an operator-valued distribution defined on
test-spinors of compact support, which would correspond to elements in
$\cD(\bR^4,V_{\rho})$. Thus, we assume here that $\Fi_0$ can be
extended to an operator-valued distribution on $\Srv$ with the above
stated properties.

Now we quote the spin-statistics theorem for a quantum field theory on
Minkowski spacetime which is proved in \cite{StrWi} for complex linear
irreducible $\rho$ and in \cite{Jost} for real linear irreducible
$\rho$. (In fact, the results in \cite{StrWi,Jost} are
slightly more general since Bosonic commutation relations are only
required in the ordinary sense there.)
\\[10pt]
{\bf Theorem B.1} {\it 
Suppose that $\FFi_{{\bf M}_0}$ is a quantum field theory on Minkowski
spacetime fulfilling the above listed conditions 1. -- 8. Then the
following two cases imply that $\Fi_0(f) = 0$, $f \in \Srv$, and
hence that $\cF_0(O) = \bC \cdot 1$ holds for all bounded open regions
$O$ in Minkowski spacetime:
\\[6pt]
($\alpha$) Bosonic commutation relations hold and the field is of
half-integer spin type ($k +l$ is odd).
\\[6pt]
($\beta$) Fermionic commutation relations hold and the field is of
integer-spin type ($k + l$ is even).
}
\\[24pt]
{\large \bf Appendix C}
\setcounter{section}{3}
\setcounter{equation}{0}
%-----------------------------------------------------------------------
\\[20pt]
In this Appendix we will explain how a generally covariant quantum
field theory over $\cG$ may be viewed as a covariant functor between
the category $\cG$ and a category $\cN$ of nets of von Neumann algebras over
manifolds (more generally, one could consider $\cN$ as the category of
isotonuos families of Neumann algebras indexed by directed index sets,
but we don't need that generality here). A similar functorial
description has been given by Dimock \cite{Dim.D}
 for the case that the morphisms
of $\cG$ are global isomorphisms, and that $\cN$ is a category of
$C^*$-algebraic nets. Here, we take the morphisms of $\cG$
to be the local isomorphisms, and correspondingly we have to consider
local morphisms for $\cN$.

We now consider $\cG$ as a category whose objects are the
four-dimensional, globally hyperbolic spacetimes with a
spin-structure.
Given ${\bf M}_1$ and ${\bf M}_2$ in $\cG$, we define the set of
morphisms hom$({\bf M}_1,{\bf M}_2)$ to consist of the local
isomorphisms between ${\bf M}_1$ and ${\bf M}_2$. We also add to
hom$({\bf M}_1,{\bf M}_2)$ a trivial morphism $\boldsymbol{0}$. (In
fact, $\boldsymbol{0}$ should be indexed by ${\bf M}_1$ and ${\bf
  M}_2$, but that is inconvenient and will be skipped as there is no
danger of confusion.) The composition of two morphisms $\TTh_a\in {\rm
  hom}({\bf M}_1,{\bf M}_2)$ and $\TTh_b \in {\rm hom}({\bf M}_2,{\bf
  M}_3)$ will be defined according to the following rules: If $\TTh_a
= \boldsymbol{0}$ or $\TTh_{b} = \boldsymbol{0}$, then $\TTh_b\TTh_a =
\boldsymbol{0}$. If both $\TTh_a$ and $\TTh_b$ are non-trivial, but
$\ell_{\rm ini}(\TTh_b) \cap \ell_{\rm fin}(\TTh_a) = \emptyset$, then
also $\TTh_b\TTh_a = \boldsymbol{0}$. Otherwise, we declare
$\TTh_b\TTh_a$ to be the local isomorphism between ${\bf M}_1$ and
${\bf M}_3$ obtained by composing the bundle maps and isometries on
their natural domains, so that $\ell_{\rm ini}(\TTh_b\TTh_a) =
 \vartheta_a^{-1}(\ell_{\rm ini}(\TTh_b) \cap \ell_{\rm
   fin}(\TTh_a))$. This is reasonable because it is not difficult to
 show that the intersection of two globally hyperbolic submanifolds of
 a globally hyperbolic spacetime yields again a globally hyperbolic
 submanifold.
 The identical bundle map gives the unit element in
 hom$({\bf M},{\bf M})$, and one can straightforwardly check that also
 the associativity of morphisms is fulfilled.

The objects of the category $\cN$ are families $\cF =\{\cF(O)\}_{O\in {\rm
    orc}(X)}$ of von Neumann algbras which are indexed by the open,
relatively compact subsets of a manifold $X$ and which are subject to
the condition of isotony (cf.\ Sec.\ 4, item ($a$)).  The morphisms in
hom$(\cF_1,\cF_2)$ are local net-isomorphisms. A local net isomorphism
is a pair $(\{\alpha_{N_i}\},\phi)$ with the following properties:
$\phi : X_1 \supset N_1 \to N_2 \subset X_2$ is a diffeomorphism
between open subsets of the manifolds $X_1$ and $X_2$ which relate to
the indexing sets of $\cF_1$ and $\cF_2$ in the obvious
manner. $\{\alpha_{N_i}\}_{N_i\in {\rm orc}(N_1)}$ is a family of von
Neumann algebraic isomorphisms $\alpha_{N_i} : \cF_1(N_i) \to
\cF_2(N_f)$ with $N_f = \phi(N_i)$ obeying the covariance property
$$ \alpha_{N_i}(\cF_1(O)) = \cF_2(\phi(O))\,,\quad O \in {\rm
  orc}(N_i)\,.$$
As before, we add to the local net-isomorphisms in hom$(\cF_1,\cF_2)$
a trivial morphism $\boldsymbol{0}$ (which may here be concretely
thought of as the map which sends each algebra element in the net $\cF_1$
to the algebraic zero element in the net $\cF_2$). The composition
rule for morphisms is then analogous as before,  we only have
to specify the case of two net-isomorphisms $({\alpha_{N_i},\phi}) \in
{\rm hom}(\cF_1,\cF_2)$ 
and $(\beta_{N'_i},\phi') \in {\rm hom}(\cF_2,\cF_3)$ when $\ell_{\rm
  ini}(\phi') \cap \ell_{\rm fin}(\phi) \ne \emptyset$. In this situation,
we define the composition of the two morphisms as the element
$(\g_{N_i},\psi)$ in
hom$(\cF_1,\cF_3)$ where $\psi$ is $\phi'\lcrc
\phi$ restricted to $\phi^{-1}(\ell_{\rm ini}(\phi') \cap \ell_{\rm
  fin}(\phi))$, and for any open, relatively compact subset $N_i$ in
$\ell_{\rm ini}(\psi)$ we define 
$$ \g_{N_i} = \b_{\phi(N_i)} \lcrc \a_{N_i}\,. $$
Again, each hom$(\cF,\cF)$ contains the identical map as an identity,
and one may check the associativity of the composition rule.

Then the covariance structure (condition (C) of Def.\ 4.1) of a
generally covariant quantum field theory is that of a covariant
functor ${\sf F}: \cG \to \cN$ which assigns to each object ${\bf M}
\in \cG$ an object ${\sf F}({\bf M}) = \{\cF(O)\}_{O \in {\rm
    orc}(M)}$ in $\cN$, and which assigns to each (non-trivial) morphism
$\TTh = (\Theta,\vartheta)$ of $\cG$ a morphism
${\sf F}(\TTh) =(\a_{\TTh,N_i},\vartheta)$ of $\cN$.
Moreover, ${\sf F}$ maps trivial morphisms to trivial morphisms.
 Diagrammatically, one has
$$
\begin{CD}
{\bf M}_1 @>{\sf F}>> \{\cF_1(O)\}_{O \in {\rm orc}(M_1)} \\
@V{\TTh}VV  @VV(\{\a_{\TTh,N_i}\},\vartheta)V\\
{\bf M}_2 @>{\sf F}>> \{\cF_2(U)\}_{U \in {\rm orc}(M_2)}
\end{CD}
$$

%%------------------------------------------------------------------------

\end{document}